\documentclass[showpacs,preprint,superscriptaddress,amsmath,amssymb]{revtex4-1}
\usepackage{graphicx}
\usepackage{dcolumn}
\usepackage{bm}
\usepackage{epsfig}
\usepackage{subfigure}
\usepackage{graphics}
\usepackage{amssymb}
\usepackage{amsthm}
\usepackage{amsmath}
\usepackage{array}
\usepackage{color}

\begin{document}

\preprint{APS/123-QED}

r\title{Phase-field-based lattice Boltzmann modeling of large-density-ratio two-phase flows}

\author{Hong Liang}
\email[Email:~]{lianghongstefanie@163.com}
 \affiliation{Department of Physics, Hangzhou Dianzi University - Hangzhou 310018, China}

\author{Jiangrong Xu}
 \affiliation{Department of Physics, Hangzhou Dianzi University - Hangzhou 310018, China}

 \author{Jiangxing Chen}
 \affiliation{Department of Physics, Hangzhou Dianzi University - Hangzhou 310018, China}

  \author{Huili Wang}
\affiliation{School of Mathematics and Statistics, Huazhong University of Science and Technology, Wuhan, 430074, China}%

 \author{Zhenhua Chai}
\affiliation{School of Mathematics and Statistics, Huazhong University of Science and Technology, Wuhan, 430074, China}%
\affiliation{Hubei Key Laboratory of Engineering Modeling and Scientific Computing, Huazhong University of Science and Technology, Wuhan 430074, China}%

\author{Baochang Shi}
\email[Email:~]{shibc@hust.edu.cn.}
\affiliation{School of Mathematics and Statistics, Huazhong University of Science and Technology, Wuhan, 430074, China}%
\affiliation{Hubei Key Laboratory of Engineering Modeling and Scientific Computing, Huazhong University of Science and Technology, Wuhan 430074, China}%

\date{\today}

\begin{abstract}
In this paper, we present a simple and accurate lattice Boltzmann (LB)
model for immiscible two-phase flows, which is able to deal with large density contrasts.
This model utilizes two LB equations, one of which is used
to solve the conservative Allen-Cahn equation, and the other is adopted to
solve the incompressible Navier-Stokes equations. A novel forcing distribution
function is elaborately designed in the LB equation for the Navier-Stokes
equations, which make it much simpler than the existing LB models. In addition,
the proposed model can achieve superior numerical accuracy compared
with previous Allen-Cahn type of LB models. Several benchmark two-phase
problems, including static droplet, layered Poiseuille flow, and Spinodal decomposition
are simulated to validate the present LB model. It is found that the
present model can achieve relatively small spurious velocity in the
LB community, and the obtained numerical results also show good
agreement with the analytical solutions or some available results.
At last, we use the present model to investigate the droplet impact on a thin
liquid film with a large density ratio of 1000 and the Reynolds
number ranging from 20 to 500. The fascinating phenomena of droplet
splashing is successfully reproduced by the present model and the
numerically predicted spreading radius exhibits to obey the power
law reported in the literature.

\end{abstract}
\maketitle

\section{Introduction}
Two-phase flows are ubiquitous in nature and engineering applications. Numerical modeling of such flows
becomes an important complement to experimental studies with the rapid development of computational science, while it
may face some certain challenges owing to complex interfacial dynamics involving multiple space and time scales.
Physically, the interfacial phenomenon can be recognized as a natural consequence of intermolecular interactions.
In this regard, the lattice Boltzmann (LB) method~\cite{Succi, Guo}, based on the mesoscopic kinetic theory, becomes a suitable candidate
to model and simulate two-phase flows.

Over the past three decades, the LB method has received great success in modeling multi-phase fluid systems~\cite{Succi, Guo} and
some nonlinear equation systems~\cite{Shi, Chai}. The reasons behind its success lie in the algorithmic simplicity, nature parallelization
and easy implementation of complex boundary. Additionally, thanks to the kinetic nature, the LB method can deal with the
intermolecular interactions in a straightforward manner, which is also regarded as its unique advantage that distinguishes
from the traditional CFD methods. Up to now, a variety of LB models for multiphase flows have been proposed, which mainly fall
into four categories, including color-gradient model~\cite{Gunstensen}, pseudo-potential model~\cite{Shan}, free energy model~\cite{Swift}
and phase-field based model~\cite{He, Lee1, Lee2, Liang1, Liang2}. For the detailed expositions, the readers can refer to the recent
reviews~\cite{Liu, Li1} on LB approaches for multiphase flows and the references therein.

Most of the previously proposed LB models are only able to handle
two-phase flows with small or moderate density contrasts. Generally,
the density ratio of liquid and vapor phases is larger than 100, and
it even could approach 1000 for a realistic water-air two-phase
system. Within this context, to develop a multiphase model that can
simulate large-density-ratio flows is an attractive topic in LB
community. Inamuro {\it{et al}.}~\cite{Inamuro} proposed a first LB model based on
the free-energy method that can tolerate large density differences.
While they needs to solve an additional Poisson equation for the
pressure to enforce the incompressible condition, which seems to be
complex, and undermines the simplicity of LB method. Besides, an
empirical cutoff value is used to determine fluid density, which
could lead to the violation of the mass conservation, like the level
set method~\cite{Sethian}. The extension of the original pseudo-potential model~\cite{Shan} to
large density ratio cases was attributed to Yuan and Schaefer~\cite{Yuan}.
They evaluated the performances of different equations of state in
the pseudo-potential model and found that large density ratio can be
reached with a suitable choice of equation of state. While it is noticed
that their studies only focus on the stationary two-phase problems, and
the model will suffer from some limitations more or less when it is readily applied
to dynamic problems. To remove this limitation, Li {\it{et al.}}~\cite{Li2}
presented an improved pseudo-potential model that can satisfy
thermodynamic consistency. Meanwhile it can improve numerical
stability of the pseudo-potential method at a large density ratio
for dynamic flows, which was demonstrated by the simulation of
droplet splashing on a liquid film with the largest density ratio of
700. But they also declared in the literature that it would induce
numerical instability when the density ratio was increased to 1000
in this case. Ba {\it{et al.}}~\cite{Ba} also developed a color-gradient-based LB
model for simulating two-phase flows with high density ratio, in
which a modified equilibrium distribution function and a simple
source term are introduced. They significantly improved the
performance of the color-gradient method and achieved satisfactory
results in the simulation of droplet splashing problem with the
largest density ratio of 100. On the other hand, several researchers have also
attempted to develop a large-density-ratio LB model based on the
phase-field theory, which has become increased popular in modeling
multiphase flows~\cite{Jacqmin}. Zheng {\it{et al}.}~\cite{Zheng} proposed a LB two-phase model
based on the Cahn-Hilliard phase field equation and claimed that
their model can simulate large-density-ratio flows. Actually, it is noted
that they only consider the Navier-Stokes equations on the average density of
binary fluids instead of the real fluid density, and therefore their
model in theory is only able to deal with density-matched binary
fluids, which is also numerically proved by Fakhari and Rahimian~\cite{Fakhari1}.
Lee {\it{et al}.}~\cite{Lee1, Lee2} also presented another LB model for
large-density-ratio two-phase flows from the phase field viewpoint.
The key point of their model in achieving large density ratio is the
use of a stable mixing difference scheme for computing gradient
terms, while it will induce the violation of mass and momentum
conservation~\cite{Lou}. Besides, the inconsistency between the recovered
interfacial equation and the target equation in their models was also found~\cite{Zu, Liang1}.
Wang {\it{et al}.}~\cite{Wang} proposed an interesting LB flux model for two-phase
flows with large density ratios, in which a stable high-order WENO
difference scheme is used to solve the Cahn-Hilliard equation, and a
like finite volume method for particle distribution function is
utilized to solve the incompressible Navier-Stokes equations. Recently, Ren et al.~\cite{Ren}
proposed a LB model from the perspective of the Allen-Cahn phase field equation,
while they only concentrated on two-phase flows limited to small or moderate density ratio,
and whether it can be applicable for large-density-ratio flows has not been discussed. Besides,
their model contains many complex gradient terms, and seems to be implemented difficultly.
More recently, Fakhari and Bolster~\cite{Fakhari2} developed a simple LB model based on
the Allen-Cahn phase field equation that can simulate large-density-ratio two-phase flows.
However, it is found that the model contains some artificial terms in the recovered interfacial
equation, which may affect the numerical accuracy~\cite{HWang,Ren}.

In this paper, we intend to present a simple, accurate and also
robust two-phase model for large-density-ratio flows in the
framework of LB method. The proposed LB model is based on the
Allen-Cahn phase field theory, which only contains a most
second-order gradient term. Therefore the present model can achieve
a lower numerical dispersion in interface tracking, compared with
the previous LB models~\cite{He, Lee1, Lee2, Fakhari1, Zu, Liang1} based on the four-order Cahn-Hilliard
equation. This is the main contribution to obtain a better numerical
stability at a large density ratio. In addition, a novel force
distribution function is introduced in this model, which can be much
simpler than those of the existing LB models~\cite{Lee1, Lee2, Zu, Liang1, Ren, Fakhari2}.
The inconsistency of the recovered interfacial equation with the target equation in
Fakhari's model~\cite{Fakhari2} is also remedied in this model by the incorporation
of a proper source term~\cite{Ren, HWang}. Through the Chapman-Enskog analyis, our model
can recover both the conservative Allen-Cahn and incompressible
Navier-Stokes equations correctly, which can be more accurate than the previous
Allen-Cahn based LB models~\cite{Ren, Fakhari2}. The rest of the paper is arranged as follows. In Sec.
\ref{method}, the macroscopic governing equations are first given, and a novel
LB model for two-phase flows based on the Allen-Cahn phase field
theory is then presented. Numerical experiments to validate the
present model and a detailed comparison with some previous LB models
can be found in Sec. \ref{sec:Results}, and finally we made a brief
summary in Sec. \ref{sec: sum}.


\section{LB model for two-phase flows}\label{method}

In this section, we first give a brief introduction on the governing
equations in the framework of the Allen-Cahn phase-field theory~\cite{Sun, Chiu}, and
then present a LB model for two-phase incompressible flows. Based on
the collision operator used, the LB method can be roughly divided
into three categories: the single-relaxation-time or so-called BGK
method~\cite{Qian}, the two-relaxation-time method~\cite{Ginzburg}, and the
multiple-relaxation-time (MRT) method~\cite{Lallemand}. Considering its simplicity
and high computational efficiency, the present model is constructed
based on the BGK collision operator and its extension to the
advanced MRT version can be conducted directly, which constitutes
one of our future research branches.

\subsection{Governing equations}
The conservative Allen-Cahn equation can be expressed by~\cite{Sun, Chiu}
\begin{equation}
{{\partial{\phi}}\over{\partial t}} + \nabla
\cdot({\phi}{\bf{u}})=\nabla\cdot\left[{M}(\nabla {\phi}-\lambda
\mathbf{n})\right],
\end{equation}
where $M$ is the mobility, $\mathbf{n}$ is the unit vector normal to
the interface,
\begin{equation}
\mathbf{n}=\frac{\nabla \phi}{|\nabla \phi|},
\end{equation}
and $\lambda$ is a function of $\phi$ defined by
\begin{equation}
\lambda=\frac{4\phi(1-\phi)}{W},
\end{equation}
where $W$ is the interface thickness, $\phi$ taking $1$ and $0$
represents the liquid and gas phase fluids, respectively, and the
interface is marked by the contour level of $\phi=0.5$. Here we
consider the incompressible two-phase flows, and the fluid velocity
$\bf{u}$ in Eq. (1) is governed by the following Navier-Stokes
equations with the force~\cite{Unverdi},
\begin{subequations}
\begin{equation}
\nabla  \cdot {\bf{u}} = 0,
\end{equation}
\begin{equation}
{{\partial ({ \rho \bf{u}})} \over {\partial t}} + \nabla \cdot
 (\rho {\bf{u}} {\bf{u}}) =  - \nabla p + \nabla  \cdot \left[
{\mu(\nabla {\bf{u}} + \nabla {{\bf{u}}^T})} \right] + {{\bf{F}}_s}
+ {\bf{G}},
\end{equation}
\end{subequations}
where $\rho$ is the fluid density, $p$ is the hydrodynamic pressure,
$\mu$ is the dynamic viscosity by $\mu=\rho\nu$, $\nu$ is the
kinematic viscosity, ${\bf{F}_s}$ is the surface tension force, and
$\bf{G}$ is the possible body force. In the literature~\cite{Kim}, there exists
several different forms of the surface tension force, here we choose
the widely used one of the potential form in the phase field
methods~\cite{Jacqmin, Zu, Liang1, Liang3},
\begin{equation}
\mathbf{F}_s={\mu_{\phi}}\nabla{\phi},
\end{equation}
where ${\mu_{\phi}}$ is the chemical potential defined by
\begin{equation}
\mu_{\phi}=4\beta\phi(\phi-1)(\phi-0.5)-k\nabla^2\phi,
\end{equation}
where $k$ and $\beta$ are physical parameters that depend on the
interface thickness and the surface tension ($\sigma$),
\begin{equation}
k=\frac{3}{2}\sigma W,~~\beta=\frac{12\sigma}{W}.
\end{equation}

\subsection{LB model for the conservative Allen-Cahn equation}
The LB evolution equation with the BGK collision operator for the
conservative Allen-Cahn equation can be written as~\cite{Shi, HWang}
\begin{equation}
{f}_i({\bf{x}} +{{\bf{c}}_i}{\delta _t},t + {\delta _t}) -
{f}_i({\bf{x}},t) = -\frac{1}{\tau_f}[{{f}_i({\bf{x}},t) -
f_i^{eq}({\bf{x}},t)}]+ {\delta _t}F_i({\bf{x}},t),
\end{equation}
where $f_i({\bf{x}},t)$ is the particle distribution function,
${\tau_f}$ is the non-dimensional relaxation time related to the
mobility, $F_i({\bf{x}},t)$ is the source term, and a simple form of
the equilibrium distribution function $f_i^{eq}({\bf{x}},t)$ is
given by
\begin{equation}
f_i^{eq} = {\omega_i}\phi(1+\frac{\mathbf{c}_i\cdot \mathbf{u}
}{c_s^2})
\end{equation}
where $c_s$ is the sound speed, $\mathbf{c}_i$ are the discrete
velocities, and ${\omega_i}$ are the weighting coefficients.
$\mathbf{c}_i$ and ${\omega_i}$ depend on the choice of the lattice
model. For the two-dimensional flows considered here, the D2Q5 or
D2Q9 lattice model can be applied in the LB algorithm for Allen-Cahn
equation. Considering the consistency with the LB algorithm for
Navier-Stokes equations, in this work we adopt the popular D2Q9
lattice model~\cite{Qian, Liang1, Liang2, Wei1}. Then the weighting
coefficients $\omega_i$ can be given by $\omega_0=4/9$,
$\omega_{1-4}=1/9$, $\omega_{5-8}=1/36$, and the discrete velocities
$\textbf{c}_i$ are
\begin{equation}
\mathbf{c}_{i}=\left\{
\begin{array}{ll}
 (0,0)c,                                                         & \textrm{ $i=0$},   \\
 (\cos [(i-1)\pi /2],\sin [(i-1)\pi /2])c,                       & \textrm{ $i=1-4$}, \\
 \sqrt{2}(\cos [(i-5)\pi /2+\pi /4],\sin [(i-5)\pi /2+\pi /4])c, & \textrm{ $i=5-8$},
\end{array}
\right.
\end{equation}
where $c=\delta_x$/$\delta_t$ is the lattice speed with $\delta_x$,
$\delta_t$ representing the grid spacing and the time increment
respectively, and $c_s=c/\sqrt{3}$. By convention, $\delta_x$ and
$\delta_t$ are set as the length and time units, i.e.,
$\delta_x=\delta_t=1$.

To recover the Allen-Cahn equation exactly with the multi-scale
analysis, the source term $F_i$ in Eq. (8) should be defined as~\cite{Ren, HWang}
\begin{equation}
F_i=(1-\frac{1}{2\tau_f})\frac{{\omega_i}\mathbf{c}_i\cdot[\partial_t(\phi
\mathbf{u})+c_s^2\lambda \mathbf{n}]}{c_s^2},
\end{equation}
where the time derivative term $\partial_t(\phi \mathbf{u})$ is
introduced to eliminate the artificial term in the recovered
equation, which is similar to the technique used in LB models~\cite{Liang1, Liang2, Liang3}
for the Cahn-Hilliard Equation. One notice that in the existing LB
model~\cite{Fakhari2} based on the Allen-Cahn theory, the term $\partial_t(\phi
\mathbf{u})$ is not included, which results in the deviation between
the recovered equation and the target equation~\cite{Ren, HWang}.

The order parameter in the present model can be computed by
\begin{equation}
\phi = \sum\limits_i {{f_i}}.
\end{equation}
The distribution of fluid density in a two-phase system physically
is consistent with that of the order parameter. To satisfy this
physical property, the fluid density should take the linear
interpolation,
\begin{equation}
\rho = \phi(\rho_l-\rho_g)+\rho_g,
\end{equation}
where $\rho_l$ and $\rho_g$ represent the densities of the liquid
and gas phases. Following the Chapman-Enskog analysis in
Ref.~\cite{HWang}, it is found that the conservative Allen-Cahn
equation can be recovered correctly from the LB equation (8) and the
mobility can be determined by
\begin{equation}
{M} =c_s^2(\tau_f-0.5)\delta t.
\end{equation}

\subsection{LB model for the Navier-Stokes Equations}
The LB equation with the BGK collision operator for the
Navier-Stokes Equations can be expressed as~\cite{Guo1, Wei2}
\begin{equation}
{g}_i({\bf{x}} +{{\bf{c}}_i}{\delta _t},t + {\delta _t}) -
{g}_i({\bf{x}},t) = -\frac{1}{\tau_g}[{{g}_i({\bf{x}},t) -
g_i^{eq}({\bf{x}},t)}]+ {\delta _t}G_i({\bf{x}},t),
\end{equation}
where ${g}_i({\bf{x}},t)$ is the distribution function for solving
the flow field, $g_i^{eq}({\bf{x}},t)$ is its corresponding
equilibrium distribution function, ${\tau_g}$ is the dimensionless
relaxation time related to the viscosity, $G_i({\bf{x}},t)$ is the
force distribution function. To satisfy the divergence-free
condition of velocity, $g_i^{eq}({\bf{x}},t)$ should be elaborately
designed as~\cite{Liang1, Liang3}
\begin{equation}
g_i^{eq}=\left\{
\begin{array}{ll}
{p \over {c_s^2}}({\omega _i} - 1) + \rho{s_i}({\bf{u}}),              & \textrm{ $i=0$},    \\
{p \over {c_s^2}}{\omega _i} + \rho{s_i}({\bf{u}}),                    & \textrm{ $i\neq0$} \\
\end{array}
\right.
\end{equation}
with
\begin{equation}
{s_i}({\bf{u}}) = {\omega_i}\left[ {{{{{\bf{c}}_i} \cdot {\bf{u}}}
\over {c_s^2}} + {{{{({{\bf{c}}_i} \cdot {\bf{u}})}^2}} \over
{2c_s^4}} - {{{\bf{u}} \cdot {\bf{u}}} \over {2c_s^2}}} \right].
\end{equation}
For the two-dimensional flows, the D2Q9 lattice model is also
adopted for flow field and the related physical coefficients
${\omega_i}$, ${{\bf{c}}_i}$ are also chosen as those of the
previous section.

Different from the previous LB models~\cite{Lee1, Lee2, Zu, Liang1, Ren, Fakhari2}, a novel force distribution function is given by
\begin{equation}
{{G}_i}=(1-\frac{1}{2\tau_g}){\omega_i}\left[{\bf{u}}\cdot\nabla\rho+\frac{\mathbf{c}_i\cdot
\textbf{F}}{c_s^2}+\frac{{\bf{u}}{\nabla
\rho}:({{\bf{c}}_i}{{\bf{c}}_i}-{c_s^2}{\textbf{I}})}{c_s^2}\right],
\end{equation}
where $\bf{F}$ is the total force,
\begin{equation}
\bf{F}=\bf{F_s}+\bf{G}.
\end{equation}
We would like to point out that the force distribution function
given in Eq. (18) can be much simpler than those of the previous
models~\cite{Lee1, Lee2, Zu, Liang1, Ren, Fakhari2}. In addition, the present LB
model with the force term (18) can recover the Navier-Stokes
equations correctly using the Chapman-Enskog analysis (see Appendix A for the detais). Substituting
Eqs. (5) and (13) into Eq. (18), one can further simplify Eq. (18)
as
\begin{equation}
{{G}_i}=(1-\frac{1}{2\tau_g}){\omega_i}\left[\frac{\mathbf{c}_i\cdot
(\mu_{\phi}\nabla\phi+\textbf{G})}{c_s^2}+\frac{(\rho_l-\rho_g){\bf{u}}{\nabla
\phi}:{{\bf{c}}_i}{{\bf{c}}_i}}{c_s^2}\right].
\end{equation}
Taking the zeroth- and the first order moments of the distribution
function ${g_i}$, the macroscopic quantities $\bf{u}$ and $p$ can be
evaluated as~\cite{Liang1, Liang3}
\begin{subequations}
\begin{equation}
\rho{\bf{u}} ={{\sum\limits_i {{{\bf{c}}_i}{g_i}} + 0.5{\delta
_t}{{\bf{F}}}}},
\end{equation}
\begin{equation}
 p = {{c_s^2} \over {(1 - {\omega
_0})}}\left[ {\sum\limits_{i \ne 0} {{g_i}}  + {{{\delta _t}} \over
2}{\bf{u}} \cdot \nabla \rho + \rho {s_0}(\textbf{u})} \right],
\end{equation}
\end{subequations}
which can be further recast as
\begin{subequations}
\begin{equation}
\rho{\bf{u}} ={{\sum\limits_i {{{\bf{c}}_i}{g_i}} + 0.5{\delta
_t}({\mu_{\phi}\nabla\phi} + {\bf{G}})}},
\end{equation}
\begin{equation}
 p = {{c_s^2} \over {(1 - {\omega
_0})}}\left[ {\sum\limits_{i \ne 0} {{g_i}}  + {{{\delta _t}} \over
2}(\rho_l-\rho_g){\bf{u}} \cdot \nabla \phi + \rho
{s_0}(\textbf{u})} \right],
\end{equation}
\end{subequations}
with the substitution of Eqs. (5) and (13). Based on the
Chapman-Enskog analysis, the fluid kinematic viscosity can be
determined by
\begin{equation}
\nu=c_s^2(\tau_g-0.5)\delta_t.
\end{equation}
In a two-phase system, the viscosity is no longer a uniform value
due to its jump at the liquid-gas interface. There are several
manners to treat the viscosity across the interface. To be smooth
across the interface, the viscosity in the diffusion-interface
methods is usually supposed to be a linear function of the order
parameter~\cite{He},
\begin{equation}
\nu=\phi(\nu_l-\nu_g)+\nu_g,
\end{equation}
where $\nu_l$ and $\nu_g$ are the kinematic viscosities of the
liquid and gas phases. In addition to Eq. (24), another popular
treatment to determine the viscosity is the inverse linear form~\cite{Lee2, Fakhari2},
\begin{equation}
\frac{1}{\nu}=\phi(\frac{1}{\nu_l}-\frac{1}{\nu_g})+\frac{1}{\nu_g}.
\end{equation}
Oftentimes, to avoid the sharp-interface limit of the phase-field
methods, a step function is also applied for the dynamic viscosity~\cite{Ren},
\begin{equation}
\mu=\left\{
\begin{array}{ll}
\mu_l,  ~~~ \phi\geq 0.5,  \\
\mu_g,  ~~~ \phi< 0.5,     \\
\end{array}
\right.
\end{equation}
where $\mu_l$ and $\mu_g$ are the dynamic viscosities of two
different phases. The scheme (26) can achieve a considerable
accuracy in tracking the interface, while similar to the
sharp-interface methods, it could be unstable when it is applied to
interfacial dynamic problems with large topology change~\cite{Anderson}. In this
work, a simple linear form as used for density is adopted, if not
specified.

For numerical iterations, the derivative terms in the model should
be discretized with suitable difference schemes. As commonly used in
LB literatures~\cite{Zu, Liang1, Liang2, Liang3}, the gradient term is computed by the
second-order isotropic central scheme,
\begin{equation}
 \nabla \phi({\bf{x}})=\sum\limits_{i \ne 0}
{\frac{\omega_i\textbf{c}_i\phi({\bf{x}} +{{\bf{c}}_k}{\delta
_t})}{c_s^2 \delta_t}}
\end{equation}
and the laplace operator is calculated by
\begin{equation}
\nabla^2\phi({\bf{x}})=\sum\limits_{i \ne 0}
{\frac{2\omega_i[\phi({\bf{x}} +{{\bf{c}}_i}{\delta
_t})-\phi({\bf{x}})]}{c_s^2 \delta_t^2}}.
\end{equation}
Occasionally, the gradient term can be computed with the
nonequilibrium part in some certain LB approaches~\cite{Chai,HWang} for the
convection-diffusion equations. Through the multiscale analysis, it
is shown that the gradient of $\phi$ and its gradient norm in this
model can be computed by the local nonequilibrium schemes~\cite{HWang}
\begin{subequations}
\begin{equation}
|\nabla \phi|=\frac{-|C|-B}{A},
\end{equation}
\begin{equation}
\nabla \phi=\frac{C}{A+\frac{B}{|\nabla \phi|}},
\end{equation}
\end{subequations}
where $A=-c_s^2\tau_f\delta_t$, $B=M\delta_t\lambda$, and
$C=\sum\limits_i\textbf{c}_i(f_i-f_i^{eq})+0.5\delta_t(\phi
\textbf{u})$. When the $|\nabla \phi|$ and $\nabla \phi$ are
computed by Eqs. (29a) and (29b), the unit normal vector
$\textbf{n}$ then can be obtained. This treatment on the gradient
term enables the collision process being implemented locally, in
addition to the computation of $\mu_\phi$, which is one of the
striking features in LB approaches. Therefore the scheme will be
adopted in our simulations, unless otherwise stated. However, we
find that the velocity actually satisfies an implicit equation, when
Eq. (29a) and (29b) are applied for the statistics of the velocity,
and for simplicity we just applied Eqs. (27) and (28) in this step.

At the end of this section, we intend to give some remarks on the
present model for two-phase flows. Firstly, the present model is
developed based on the conservative Allen-Cahn Equation, which
contains a lower-order diffusion term compared with the fourth-order
Cahn-Hilliard equation. Therefore the Allen-Cahn based model in
theory can achieve a lower numerical dispersion in solving the index
function $\phi$ and also the density field via Eq. (13) than the
Cahn-Hilliard type of LB models. The lower-dispersion solution of
$\phi$ plays a significant role in simulating large density ratio
flows since a small deviation could be more likely to lead a
unphysically negative value of fluid density, causing numerical
instabilities. It is worth noting that a type of large-density-ratio
LB models~\cite{Lee1, Lee2} have been proposed based on the Cahn-Hilliard equation,
which is attributed to the use of a mixed scheme that combines the
central and biased differences. However, it can lead to the
violations of mass and momentum conservation~\cite{Lou}. On the contrary, in
our model the isotropic central scheme and the local nonequilibrium
scheme are applied, which not only preserve a secondary-order
accuracy in space, but also can ensure the global mass conservation
of a two-phase system. Secondly, a novel force distribution function
for flow field is proposed in the present model, which can be much
simpler than those of the existing Allen-Cahn based LB models~\cite{Ren, Fakhari2}.
It is noted that our model only contains one type of non-local gradient
term for the order parameter, which is much smaller than those of
the previous model~\cite{Ren}. In addition, the gradient term and its Laplacian
in our model can be computed with local nonequilibrium schemes,
which makes the collision process to be conducted locally if
$\mu_\phi$ has been given. While in the previous models~\cite{Ren, Fakhari2}, only the
central difference schemes are applied, and thus the collision process
cannot be conducted locally. Thirdly, both the conservative Allen-Cahn
Equation and the incompressible Navier-Stokes equations can be recovered exactly
from the present model with the multi-scale analysis. While the model of
Fakhari {\it{et al.}}~\cite{Fakhari2} contains some artificial terms in the recovered
interfacial equation. The numerical experiments conducted below will
demonstrate that the present model is more accurate than the
previous Allen-Cahn based LB models~\cite{Ren, Fakhari2}. At last, we would like to stress
that the present model is a standard LB scheme for simulating large-density-ratio
two-phase flows without the use of an advanced finite difference or finite volume method~\cite{Wang},
therefore it can naturally inherit the advantages of LB method in dealing with complex
physical boundary and parallel computing.

\section{Numerical Results and discussions}\label{sec:Results}
In this section, several typical benchmark problems, including
static droplet, layered Poiseuille flows, and Spinodal decomposition
are used to validate the present LB model for large density ratio
flows. We attempt to conduct a detailed comparison between the
present results and the analytical solutions or some available
results. At last, we also investigated droplet impact on a thin
liquid film, where the effect of the Reynolds number are discussed
in detail.

\subsection{Static droplet}
The static droplet is a basic two-phase problem, which has been
widely used to verify the developed numerical methods~\cite{Lee1, Fakhari1, Zu, Liang1, Ba, Li2}. In this
subsection, we will simulate this problem with large density ratio
to validate the present LB model. Initially, a liquid droplet with
the radius of $R=50$ surrounded by the gas phase is located at the
center of the square domain $N_y\times N_x=200\times200$ and the
periodic boundary conditions are applied at all boundaries. The
distribution profile of the order parameter is initialized by
\begin{equation}
 \phi(x,y)= 0.5+0.5\tanh \frac{2\left[R-(x-100)^2-(y-100)^2\right]}{W},
\end{equation}
which makes its value to be smooth across the interface. In the
simulation, we set the density ratio to be $\rho_l/\rho_g=1000:1$,
and some other physical parameters are given as $\nu_l=\nu_g=0.1$,
$\sigma=0.001$, $W=5$. Fig. 1(a) depicts the interface pattern of
the droplet at the equilibrium state, together with the initial one
given by Eq. (30). It can be found that they line up over each other
exactly, which indicates that the present model has a high accuracy
in track the interface. Furthermore, we quantitatively plotted in
Fig. 1(b) the density distributions along the horizontal center line
with different values of the mobility $M$. It is shown that
numerical predictions of the density field are all in good agreement
with the analytical solution.

The spurious velocity around the interface is a commonly concerned
problem in LB approaches for two-phase flows, and cannot be
completely eliminated in the framework of the LB method~\cite{Guo2}. In Fig.
1(a), we also display the velocity distribution in the whole
computational domain obtained by the present model. It can be found
that the spurious velocities indeed exist at the vicinity of the
interface, and their maximum magnitude computed by
$|\textbf{u}|_{max}=(\sqrt{u^2+v^2})_{max}$ is about
$1.6\times10^{-9}$. We also compared the present model with some
previously improved LB models in terms of the spurious velocity. It
is shown that the maximum amplitude of spurious velocities in an
improved Shan-Chen model~\cite{Yu} has the order of $10^{-3}$. Recently, Ba
{\it{et al.}}~\cite{Ba} developed an improved color-gradient based model for
high density ratio, which produced spurious velocities with the
order of $10^{-5}$. As for the Cahn-Hilliard type of LB model~\cite{Liang1}, they
can obtain spurious velocities at the level of $10^{-6}$. Therefore
we can conclude that the present LB model is able to produce
relatively small spurious velocities.

\begin{figure}
\centering
\includegraphics[width=3.2in,height=3.1in]{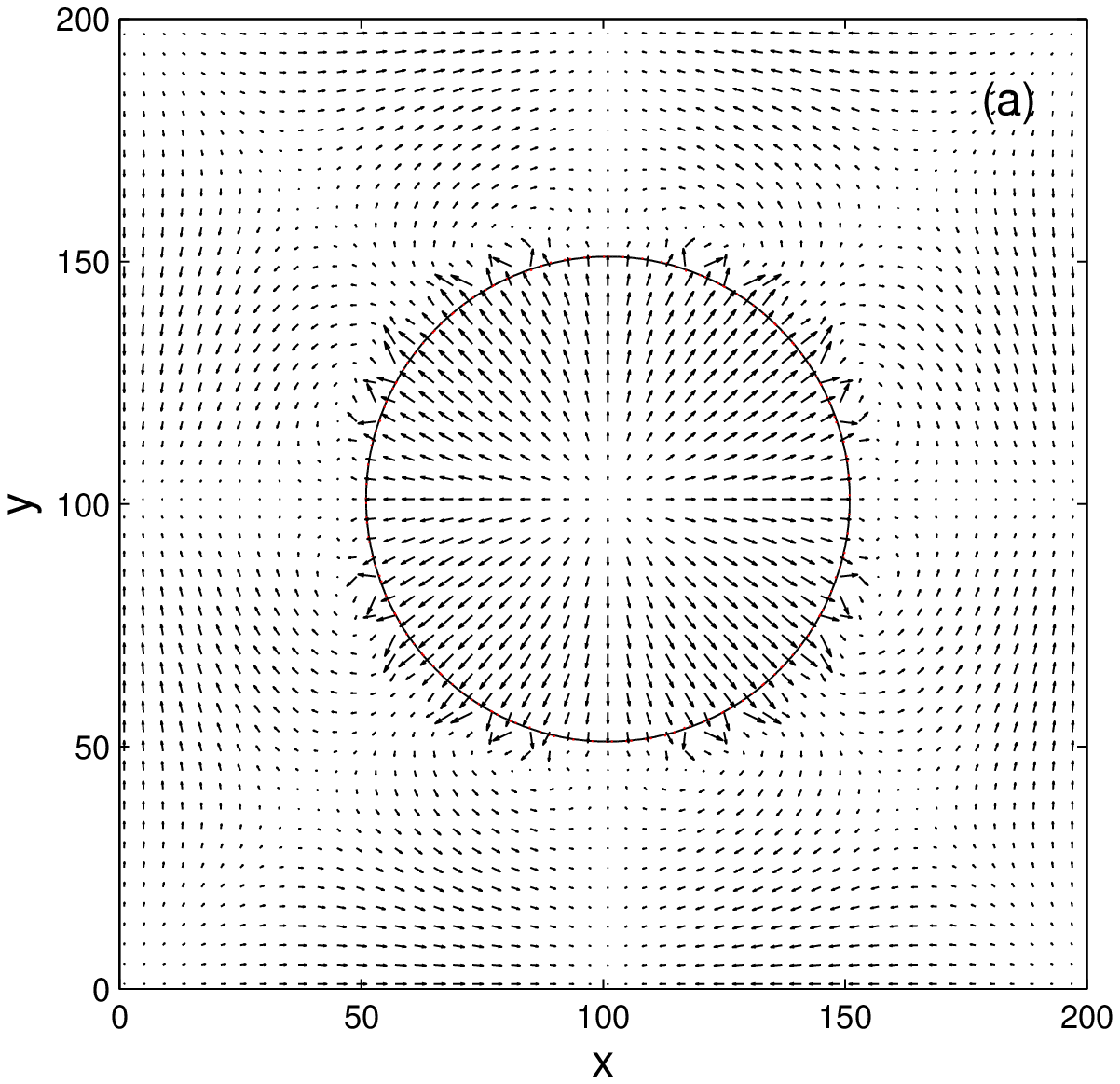}~~
\includegraphics[width=3.3in,height=3.1in]{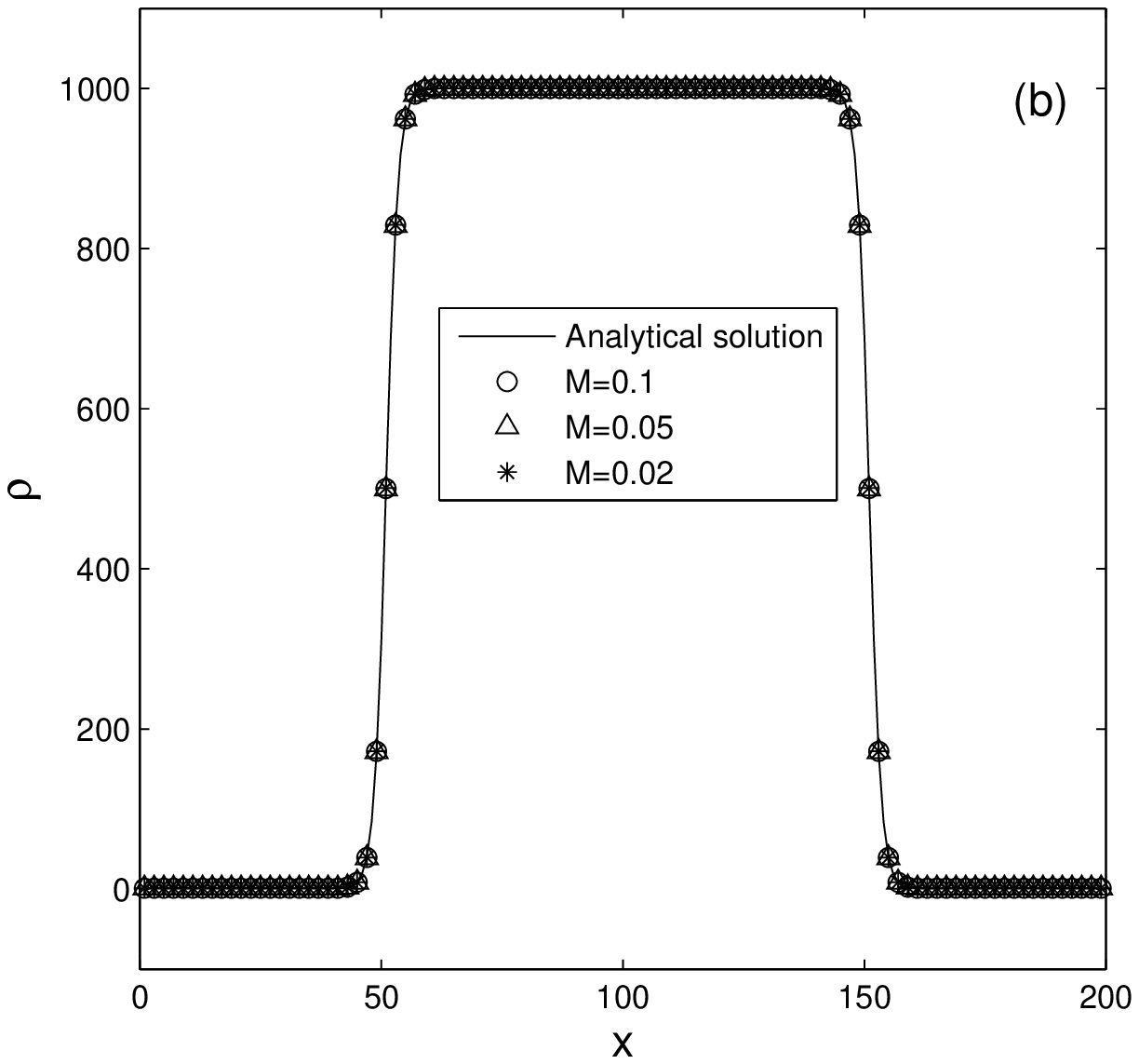}
 \tiny\caption{Static droplet tests with density ratio $\rho_l/\rho_g=1000:1$, (a) the velocity distribution of
 the whole domain at the equilibrium state. The solid and dashed lines represent the equilibrium shape of the
droplet and its initial shape, respectively; (b) the density
distributions along the horizontal
 center line $(y=N_y/2)$ at different values of the mobility.}
\end{figure}

\subsection{Layered Poiseuille flow}

\begin{figure}
\centering
\includegraphics[width=3.2in,height=2.85in]{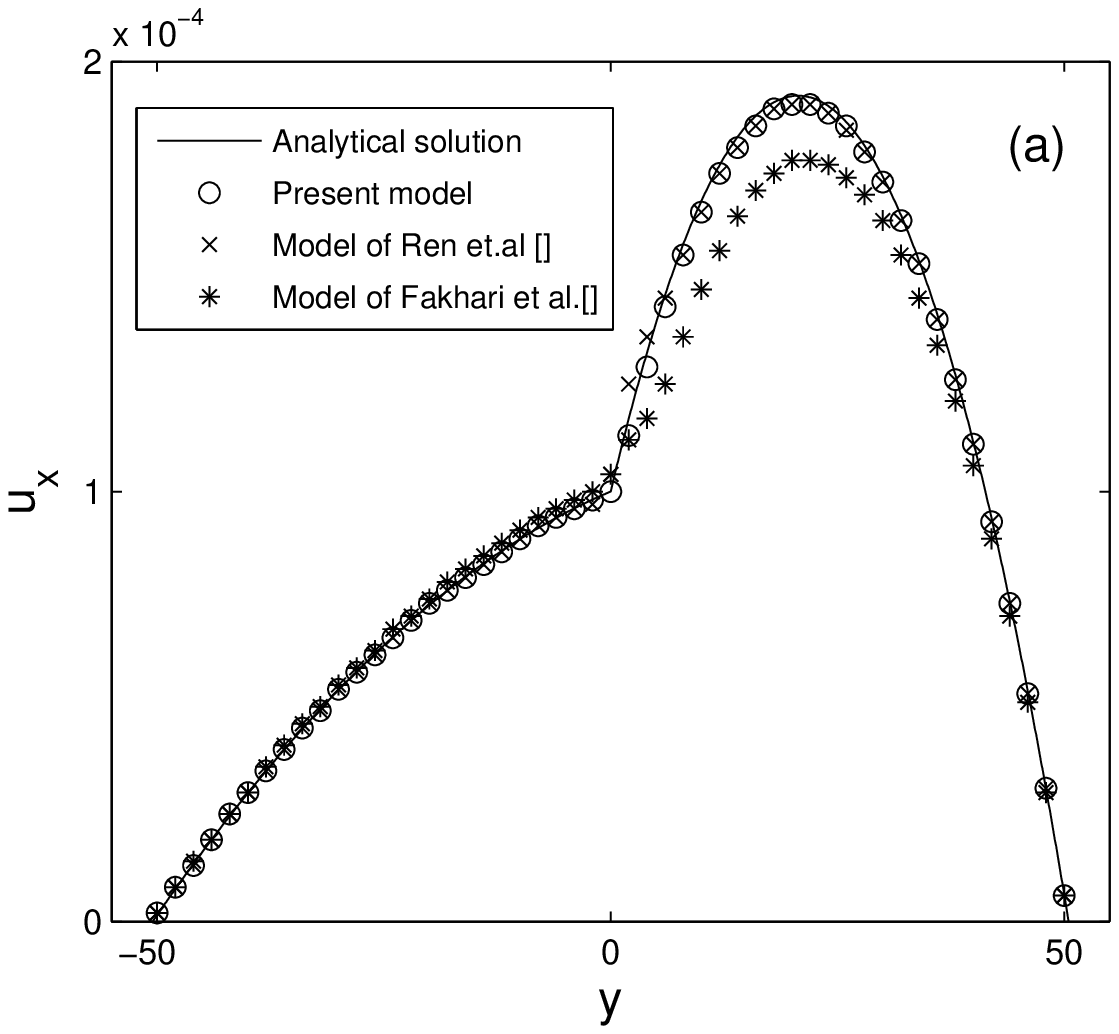}
\includegraphics[width=3.2in,height=2.85in]{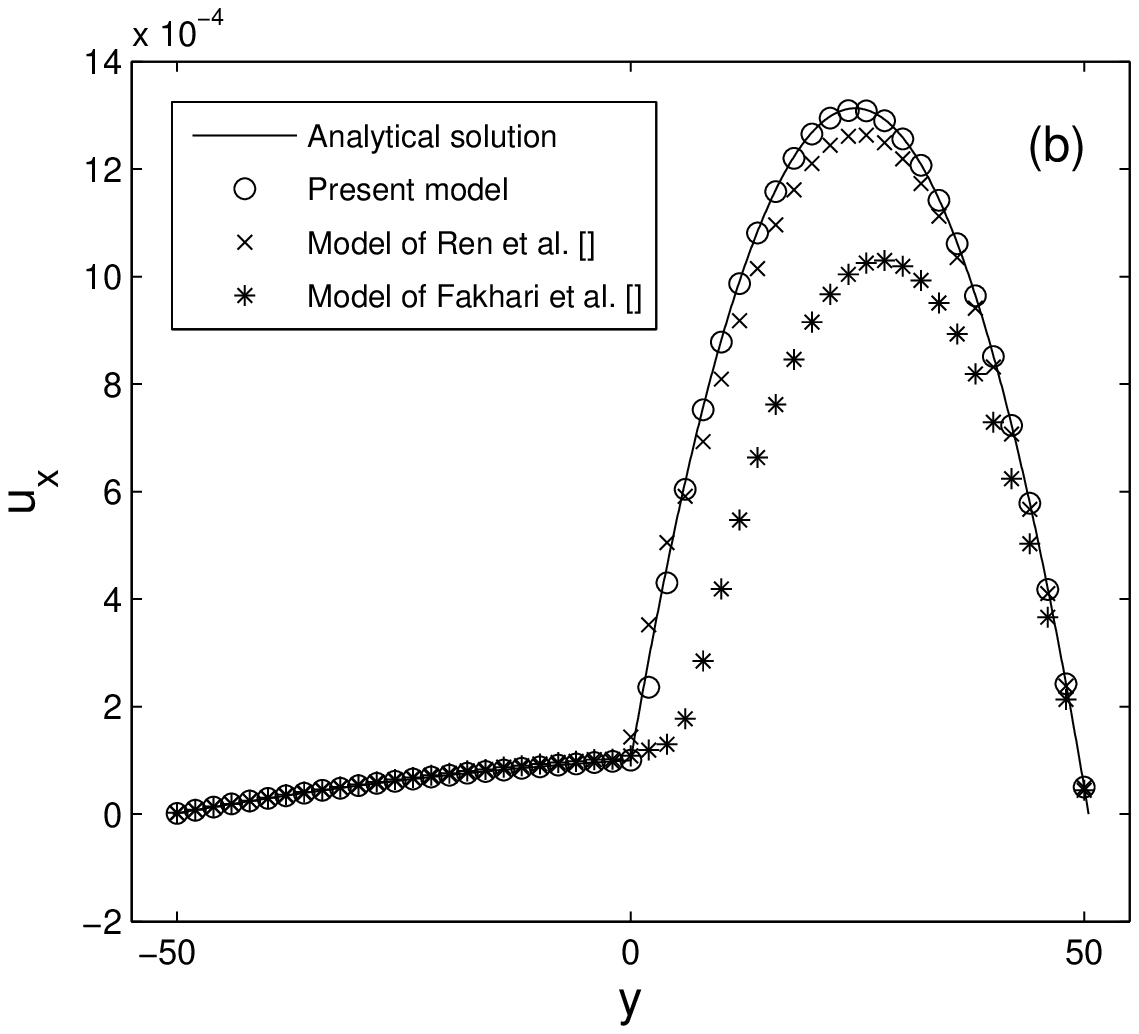}
\includegraphics[width=3.2in,height=2.85in]{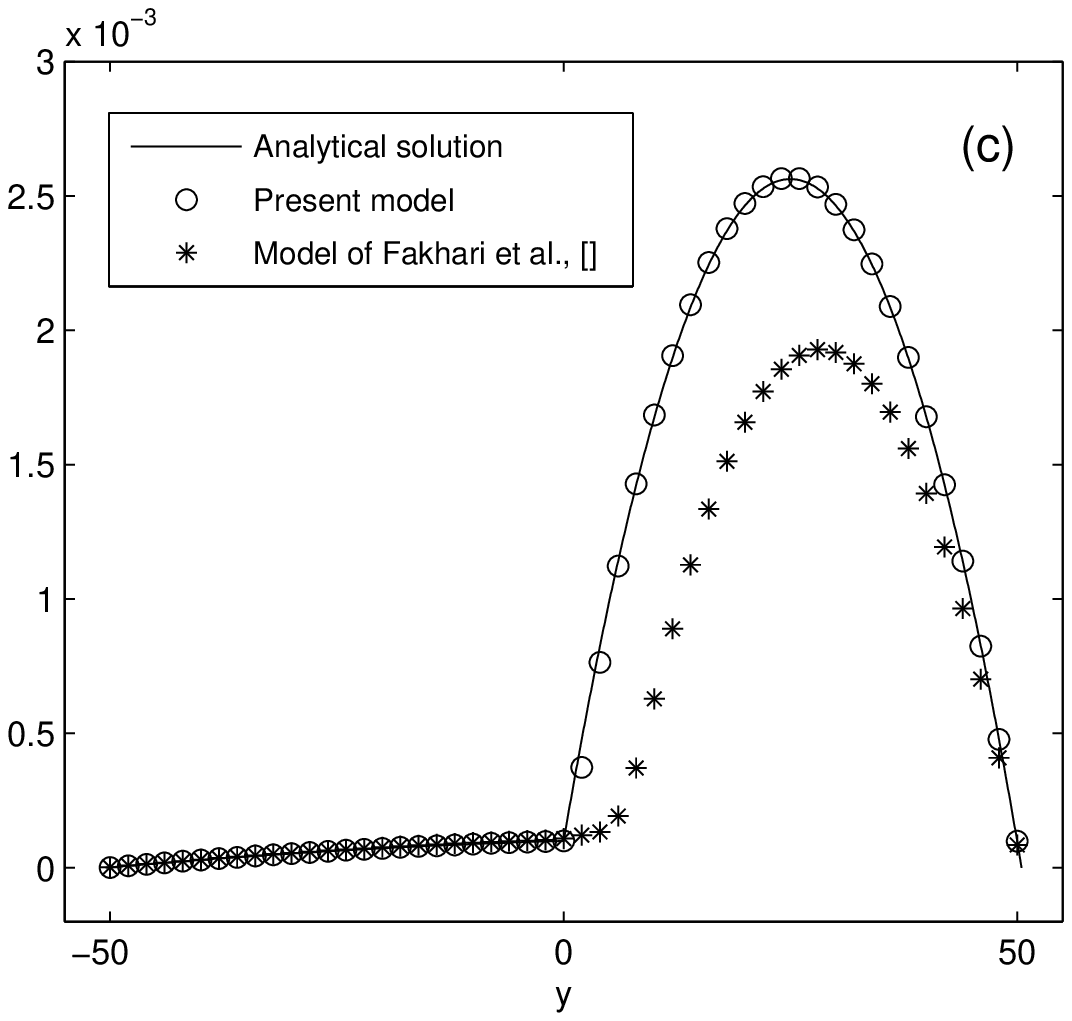}
\includegraphics[width=3.2in,height=2.85in]{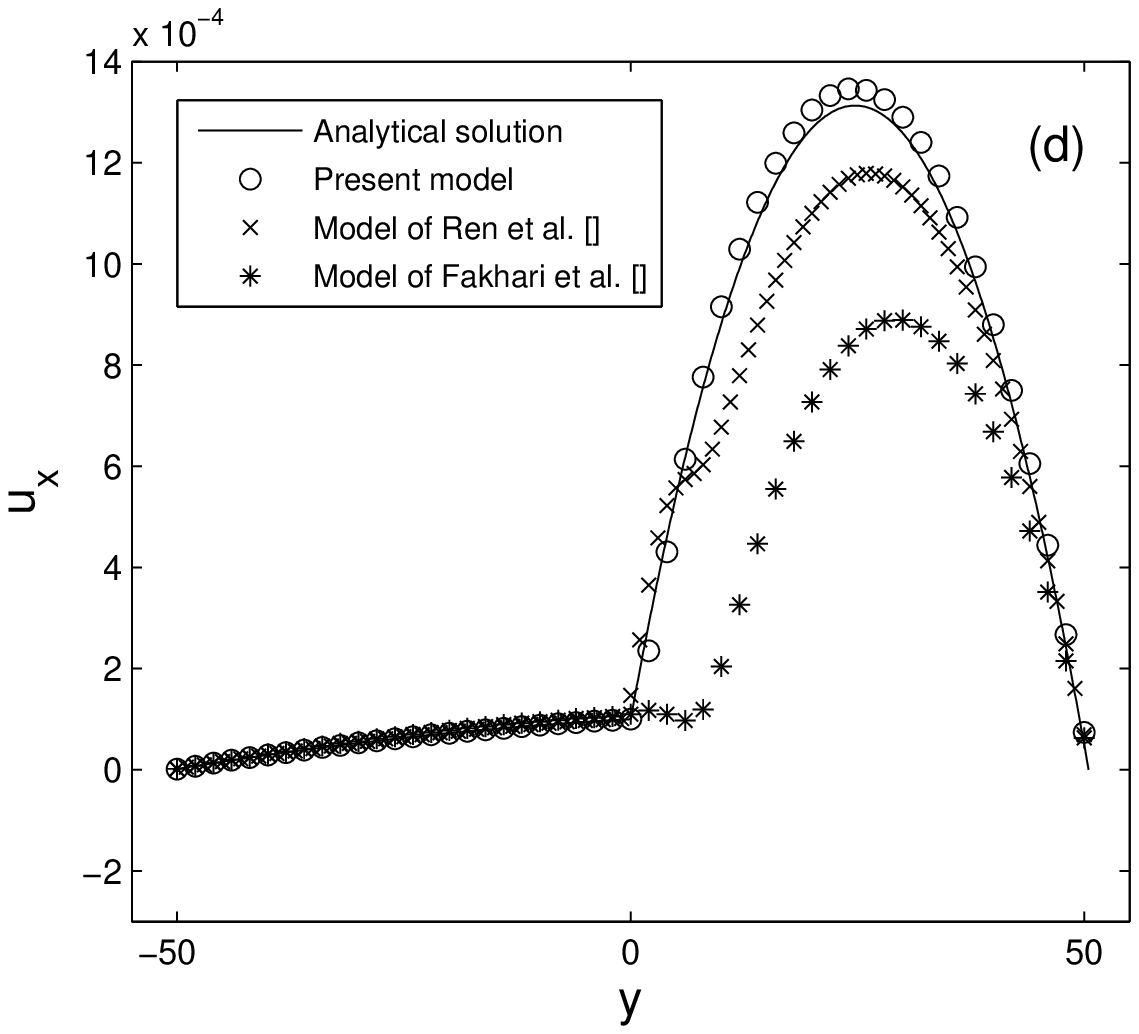}
 \tiny\caption{Comparisons of the horizontal velocity profiles obtained by the present model and the existing Allen-Cahn based LB models~\cite{Ren, Fakhari2}
 with various density ratios, (a) $\rho_l:\rho_g=10:1$; (b) $\rho_l:\rho_g=100:1$; (c) $\rho_l:\rho_g=150:1$; (d) $\rho_l:\rho_g=1000:1$.
 The solid lines represent the corresponding analytical solutions. }
\end{figure}

The layered Poiseuille flow is a classical two-phase problem, which
can provide a good benchmark for validating the developed LB
approaches~\cite{Zu, Ba,  Wang,  Ren,  Huang, Liang4}. While to our knowledge, most of the previous studies are
limited to the small density ratio less than 10, due to the
numerical instability problem. In this subsection, we will simulate
the layered two-phase Poiseuille flows with the largest density ratio of 1000,
and also conduct comparisons of the present model with the existing
Allen-Cahn based LB models~\cite{Ren, Fakhari2}. Consider a channel flow of two
immiscible fluids driven by a constant body force $\textbf{G}=(G_x,
0)$. Initially, the gas phase fluid is placed in the upper region of
$0<y\leq h$ and the region of $-h\leq y \leq 0$ is fill with the
liquid phase fluid. Periodic boundary conditions are applied in the
$x$-direction, and the bottom and top boundaries are the solid
walls, which are treated by the halfway bounce-back boundary
conditions. Based on these boundary conditions, one can derive the
analytical solution for the horizontal velocity profile ($u_x$),
\begin{equation}
 u_x(y)=\left\{
\begin{array}{ll}
\frac{G_x h^2}{2\mu_g}\left[-(\frac{y}{h})^2 - \frac{y}{h}(\frac{\mu_g-\mu_l}{\mu_g+\mu_l})+\frac{2\mu_g}{\mu_g+\mu_l} \right],   & \textrm{ $0< y\leq h$},    \\
\frac{G_x h^2}{2\mu_l}\left[-(\frac{y}{h})^2 - \frac{y}{h}(\frac{\mu_g-\mu_l}{\mu_g+\mu_l})+\frac{2\mu_l}{\mu_g+\mu_l} \right],   & \textrm{ $-h\leq y\leq 0$}, \\
\end{array}
\right.
\end{equation}
where $G_x=u_c(\mu_l+\mu_g)/h^2$, which provides a steady horizontal
velocity of $u_c$ at the center. To quantitatively describe the
accuracy of the present model and also conveniently compare with
the existing LB models, the following relative error is used,
\begin{equation}
E_{u}=\frac{\sum\limits_y
|u_x^n(y,t)-u_x^a(y)|}{\sum\limits_y|u_x^a(y)|},
\end{equation}
where the subscripts $n$ and $a$ denote the numerical and analytical
solutions.

In the simulation, the computational grid is chosen to be $N_y\times
N_x=100\times10$, and the initial distribution of the order
parameter is set as
\begin{equation}
 \phi(x,y)= 0.5+0.5\tanh \frac{2\left(0.5N_y-y\right)}{W},
\end{equation}
which gives the profile of the planar interface. $u_c$ is fixed as a
small value of $10^{-4}$, which ensures that the incompressible
limit can be satisfied, and some other related parameters are given
as $W=5$, $\sigma=0.001$, $\nu_l=0.1$ and $M=0.1$. Four different
cases of the density ratios $\rho_l/\rho_g=10,~100,~150,~1000$ are
considered, where the corresponding kinematic viscosity ratios
$\nu_g/\nu_l$ are 1 for the former three cases, and $10$ for the
latter case. Here the dynamic viscosity is given by Eq. (26) in the
present test of our model. Note that the two-phase system with
$\rho_l/\rho_g=1000$ and $\mu_l/\mu_g=100$ considered here is very
close to the realistic water-air system at the room temperature and
normal atmospheric pressure. Figure 2 shows the profiles of the
horizontal velocity ($u_x$) with various density ratios obtained by
the present model, together with the corresponding analytical
solutions. For comparisons, we also simulated the above cases with
the previous Allen-Cahn based LB models~\cite{Ren, Fakhari2} under
identical computational conditions, and the obtained numerical
results are also presented in Fig. 2. It can be observed from Fig. 2
that numerical results of the present model agree well with the
analytical solutions for all density ratios, while some obvious
discrepancies with the analytical solutions are found in the results
of the existing Allen-Cahn based LB models~\cite{Ren, Fakhari2},
especially at high density ratios. We also conducted a quantitative
comparison between the present model and the previous LB
models~\cite{Ren, Fakhari2}. The relative errors of the velocity
$u_x$ with these LB models were measured and the results are
summarized in Table {\ref{table1}. It is found that the previous
models produce large relative errors, and they all increase
significantly with the density ratio. In contrast, a much smaller
relative error can be derived by the the present model, which also
seems to be independent of the density ratio. Based on the above
discussion, we can see that the present model is more accurate than
the previous Allen-Cahn based LB models~\cite{Ren, Fakhari2}.

\begin{table*}
\caption{\label{table1}Relative errors of the horizontal velocity
$(u_x)$ in layered Poiseuille tests.}
\begin{tabular}{cccccc}
\hline
Density ratio ~~~~~~& Present LB model ~~~~~~& Model of Ren {\it{et al.}}~\cite{Ren}, ~~~~~~& Model of Fakhari {\it{et al.}}~\cite{Fakhari2} \\
($\rho_l/\rho_g$)\\
 \hline
$10$   & $8.9\times10^{-3}$ & $1.0\times10^{-2}$  & $6.2\times10^{-2}$\\
$100$  & $6.9\times10^{-3}$ & $4.4\times10^{-2}$  & $2.7\times10^{-1}$\\
$150$  & $5.4\times10^{-3}$ & $8.2\times10^{-2}$  & $3.0\times10^{-1}$\\
$1000$ & $3.2\times10^{-2}$ & $1.1\times10^{-1}$  & $3.9\times10^{-1}$\\
\hline
\end{tabular}
\end{table*}

\subsection{Spinodal decomposition}

\begin{figure}
\centering
\includegraphics[width=1.5in,height=1.58in]{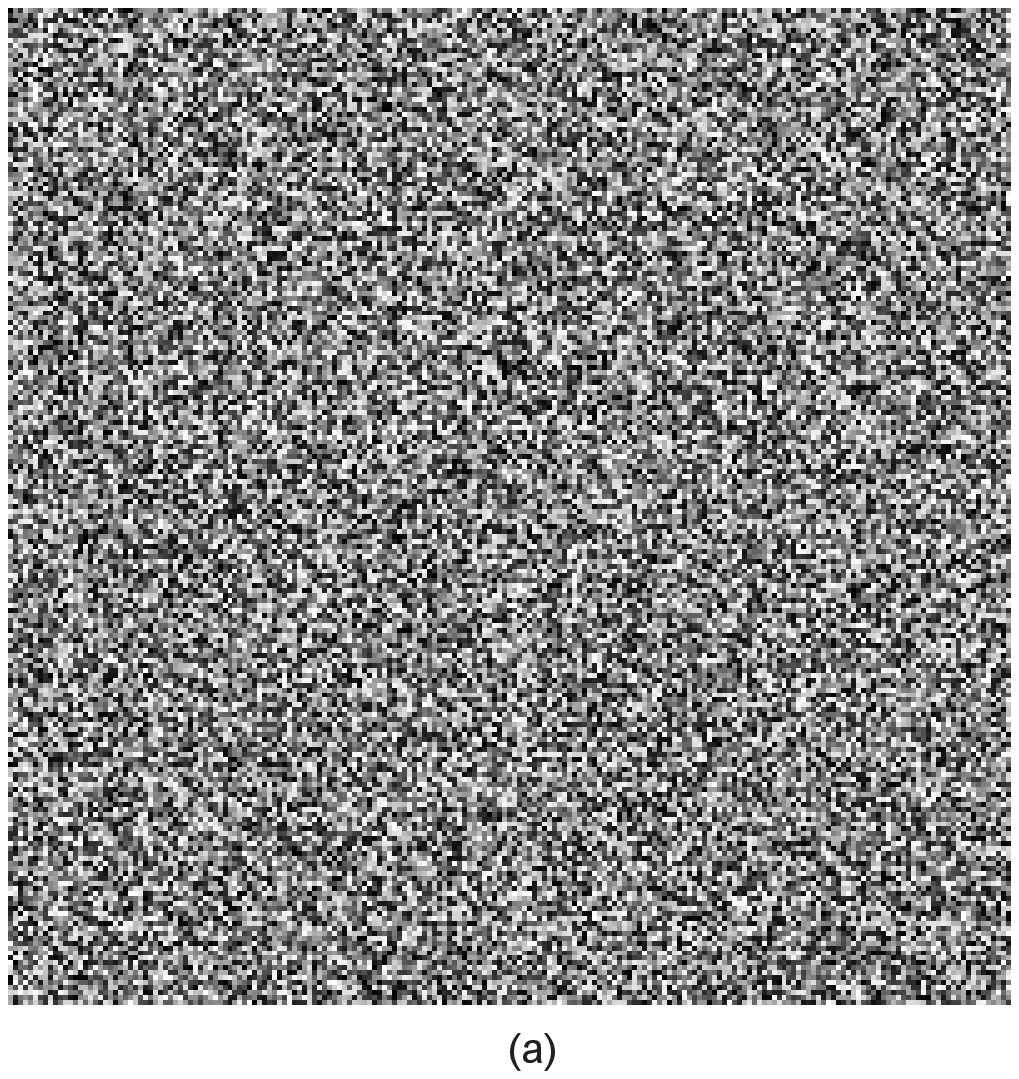}
\includegraphics[width=1.5in,height=1.58in]{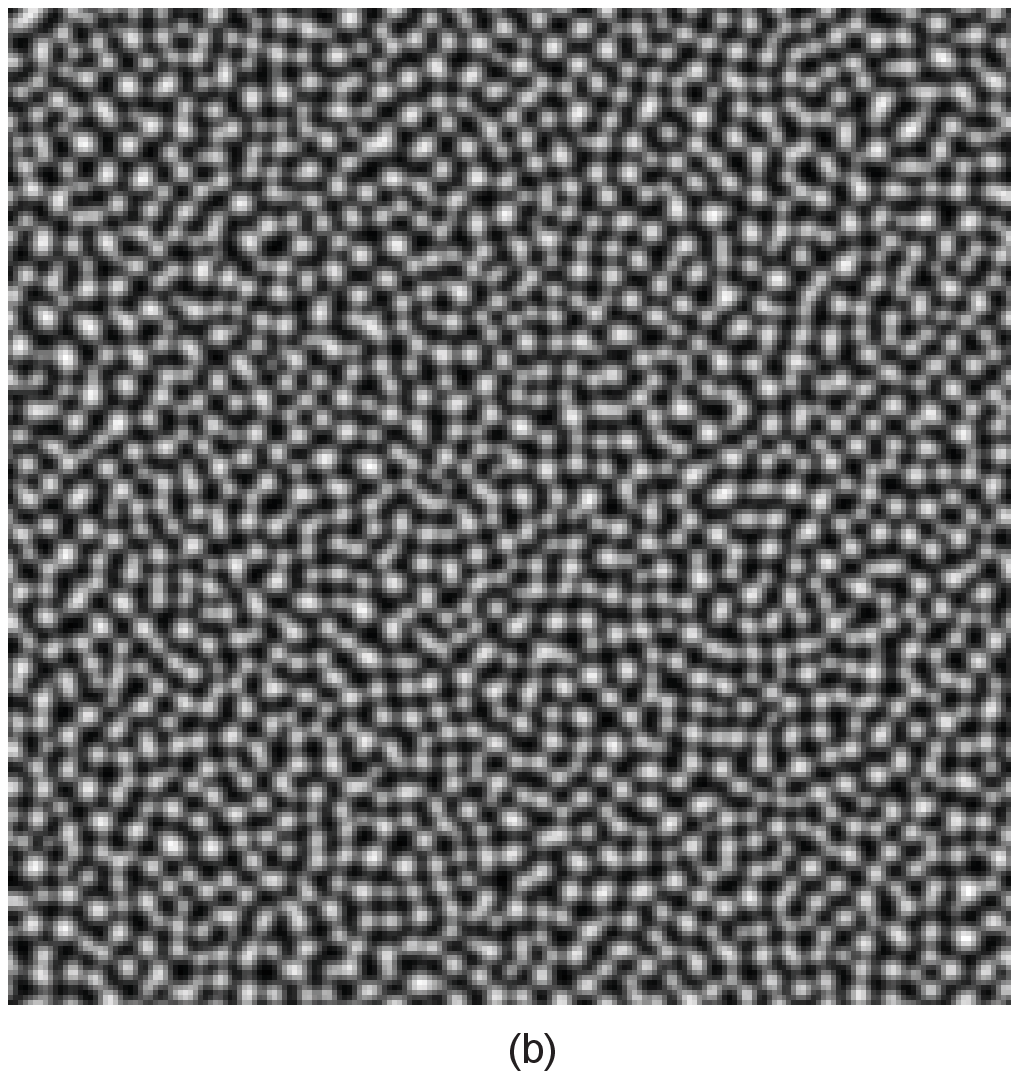}
\includegraphics[width=1.5in,height=1.58in]{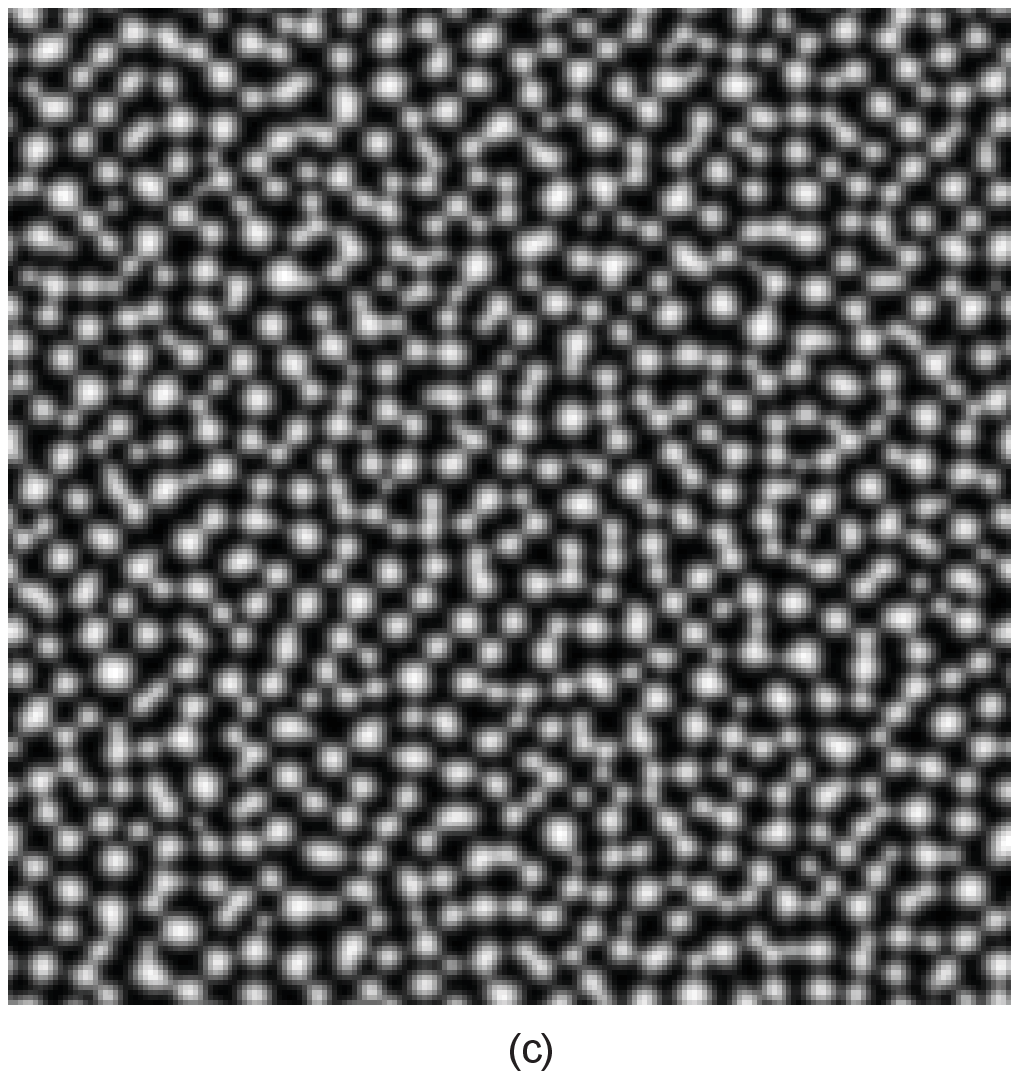}
\includegraphics[width=1.5in,height=1.58in]{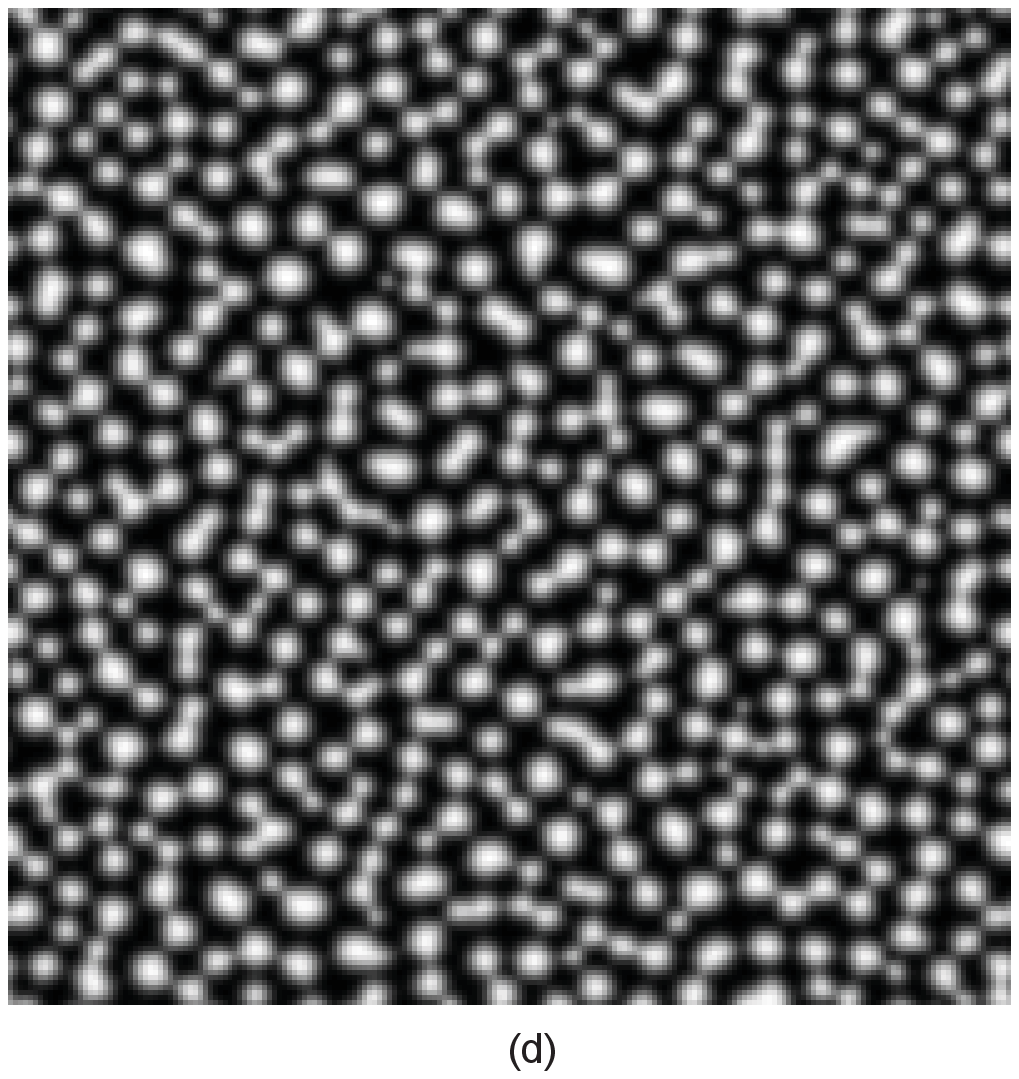}\\
\includegraphics[width=1.5in,height=1.58in]{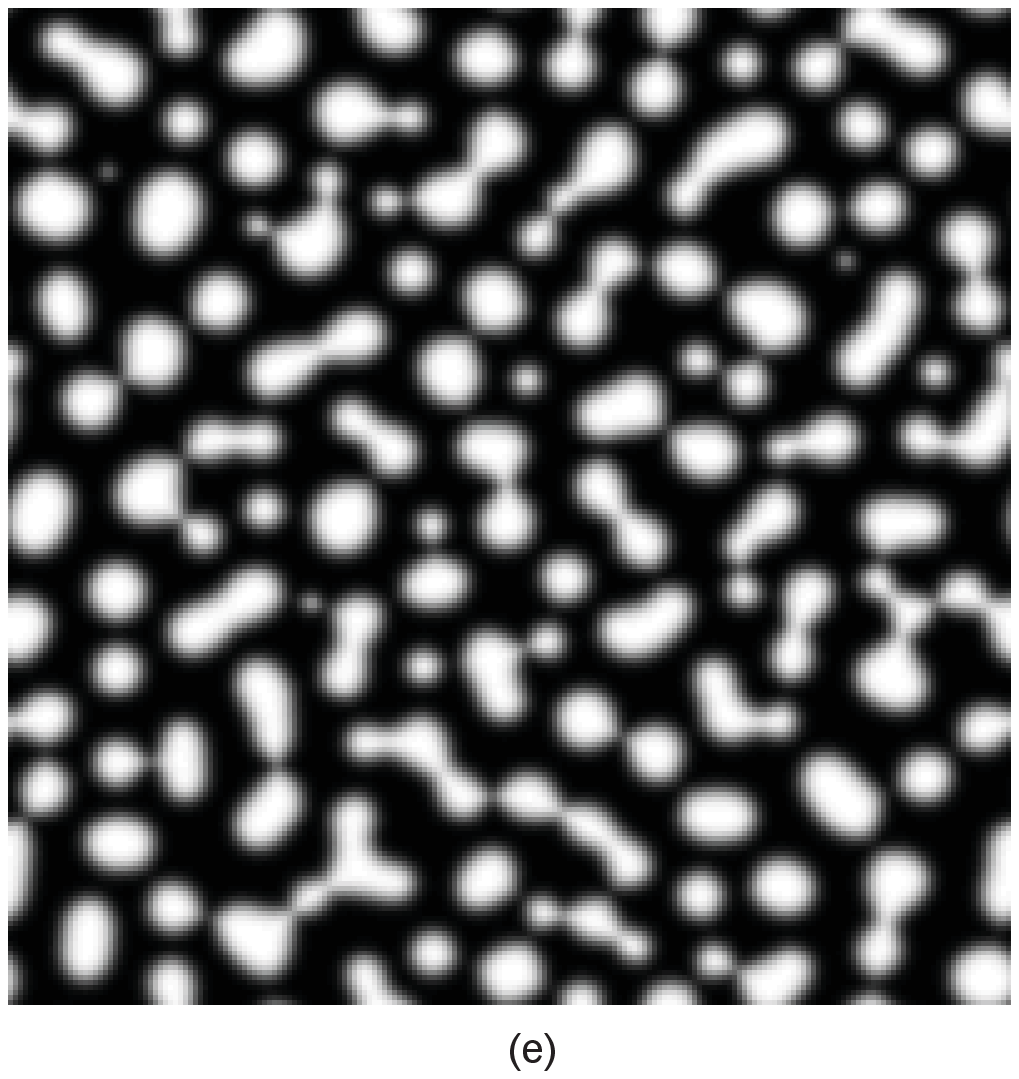}
\includegraphics[width=1.5in,height=1.58in]{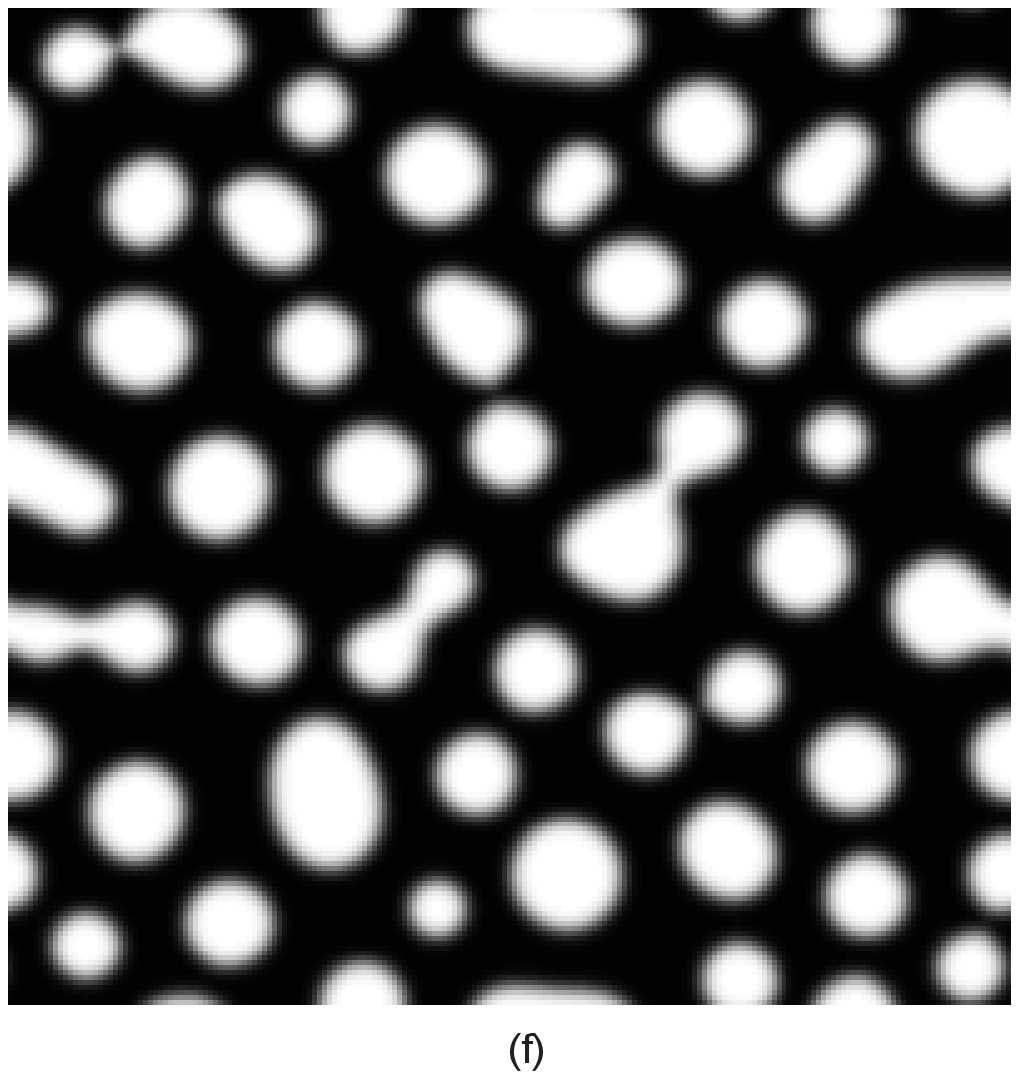}
\includegraphics[width=1.5in,height=1.58in]{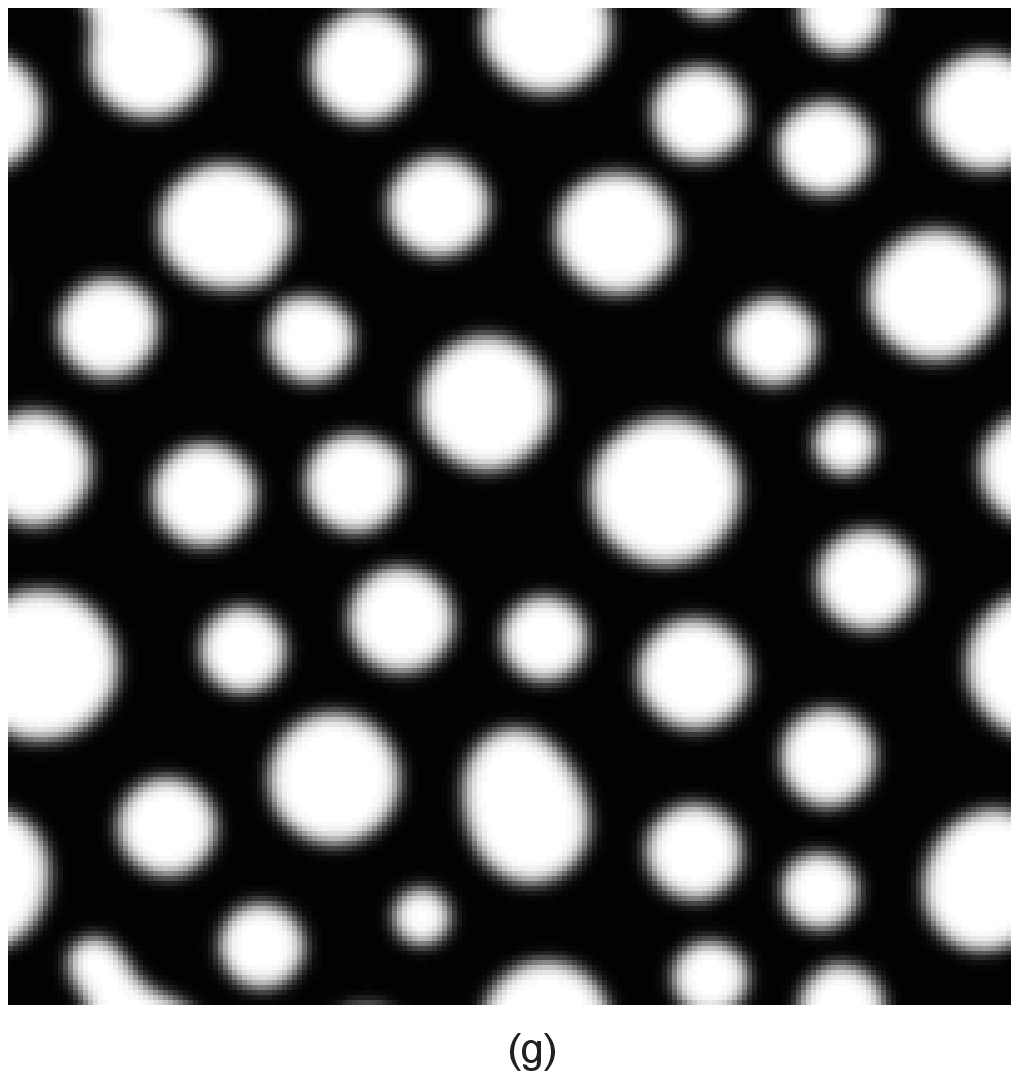}
\includegraphics[width=1.5in,height=1.58in]{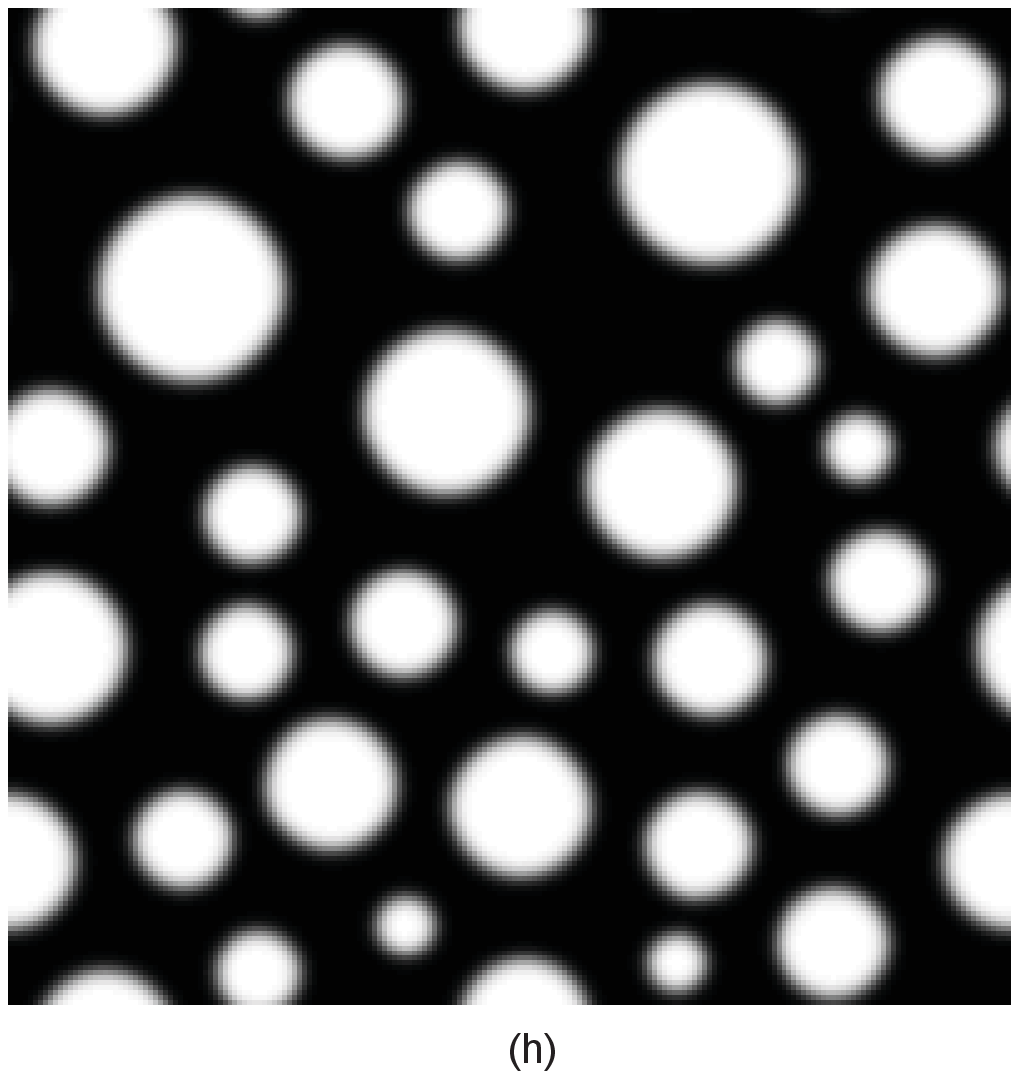}
 \tiny\caption{Time evolution of the density distribution during the phase separating processes, (a) t=0; (b) t=0.05; (c) t=0.25; (d) t=0.5; (e) t=5; (f) t=25; (g) t=50; (h) t=125.}
\end{figure}

Spinodal decomposition~\cite{Cahn} is a fundamental property of a fluid mixture
with two different species. For suitable compositions and quenches,
the initial homogeneous mixture is unstable in the presence of small
fluctuations, and then the spinodal decomposition phenomenon will
take place. This phenomenon, ubiquitous in physics and chemistry,
has been studied extensively. Several researchers have also
investigated the spinodal decomposition problem using the LB
approaches~\cite{Shan, Swift, Zu, Gan}, while they mainly focus on the process of phase
separation with small or moderate density ratios. In this
subsection, we intend to simulate this problem with the large
density ratio of 1000 by the present LB model, where the gradient
terms are computed by Eqs. (27) and (28). This exercise is devoted
to the demonstration of the capability of our method in studying
complex high-density-ratio two-phase flows. The computational mesh
used here is chosen to be $N_y\times N_x=200\times200$. The periodic
boundary conditions are applied at all boundaries. In the
simulation, the initial distribution of the order parameter with
small fluctuations can be given by
\begin{equation}
\phi(x,y)=\frac{1}{3}+\texttt{rand}(x,y),
\end{equation}
where $\texttt{rand}(x,y)$ is a random function with the maximum
amplitude of 0.01. Then a small perturbation can be imposed on a
homogeneous density field via Eq. (13), where $\rho_l$ and $\rho_g$
are set to be 1000 and 1. We only consider binary fluids with the
viscosity ratio of $\nu_g/\nu_l=10$, which approaches that of a
water-air system. The remaining parameters in the simulation are fixed as
$\sigma=0.2$, $W=4$ and $M=0.1$. Figure 3 depicts the time evolution
of the density distribution during the phase separating process,
where the time ($t$) has been nondimensionalized by the viscous time
of the liquid phase $\rho_l \nu_l W/\sigma$. It can be found that
the early stage of phase separation induces small fluctuations of
the density into large scale inhomogeneties. Then some tiny droplets
with random shapes are formed in the system. The droplet sizes
increase with time, and some of them also coalesce into the larger
ones, which leads to the eventual separation of binary fluid
components. The above phase separating processes are results of the
hydrodynamics and surface tension action, which conform to the
expectation.

\subsection{Droplet impact on a thin liquid film}

\begin{figure}
\centering
\includegraphics[width=5.5in,height=2.05in]{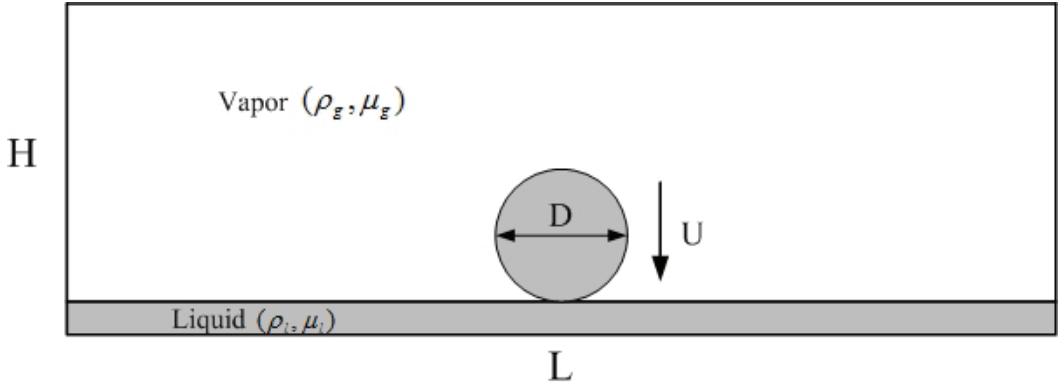}
 \tiny\caption{Schematic of the initial setup for the droplet impact on a thin liquid film.}
\end{figure}

At last, to show the capacity of the present model, we consider a
complex problem of droplet impact dynamics with large density ratio.
Droplet impact on liquid surfaces~\cite{Yarin} is a familiar spectacle in natural
event of falling raindrop on the wet ground or puddle. Further, it
plays a prominent role in many technical applications, such as ink
jet printing, spray cooling and and coating. In spite of its
ubiquity and extensive researches~\cite{Yarin, Coppola, JLee, Thoraval}, numerical simulation of such
flows still poses some challenges due to complex interfacial changes
in topology, and yet there exists a large density difference for a
water-air system. In addition, a numerical singularity may be
produced at the impact point. In this section, we will simulate a
two-dimensional droplet impact on a preexisting thin liquid film
with a large density ratio of 1000 by the present LB model, in the
absence of the gravitational field.

\begin{figure}
\centering
\includegraphics[width=5.5in,height=1.4in]{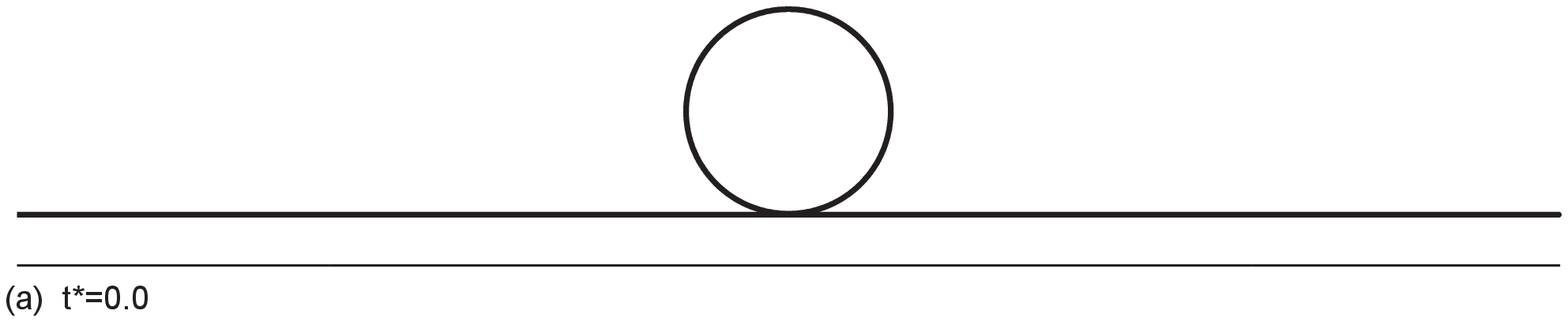}
\includegraphics[width=5.5in,height=1.4in]{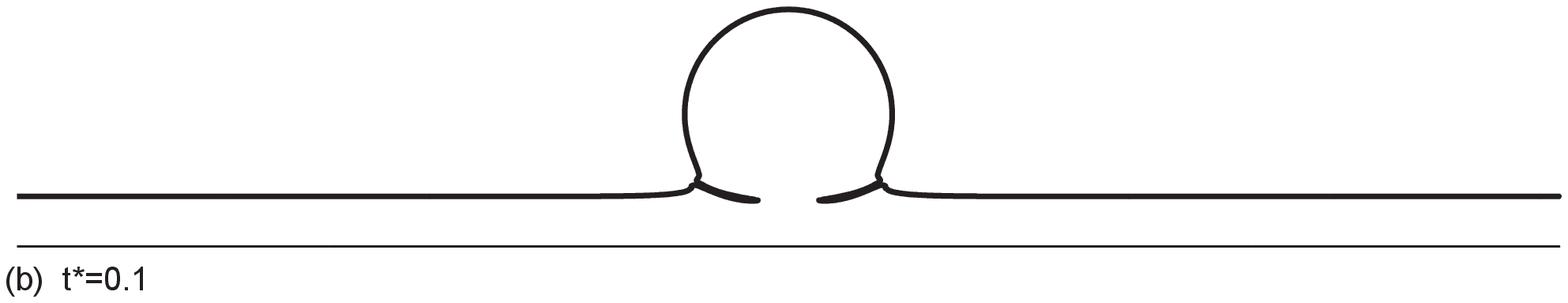}
\includegraphics[width=5.5in,height=1.4in]{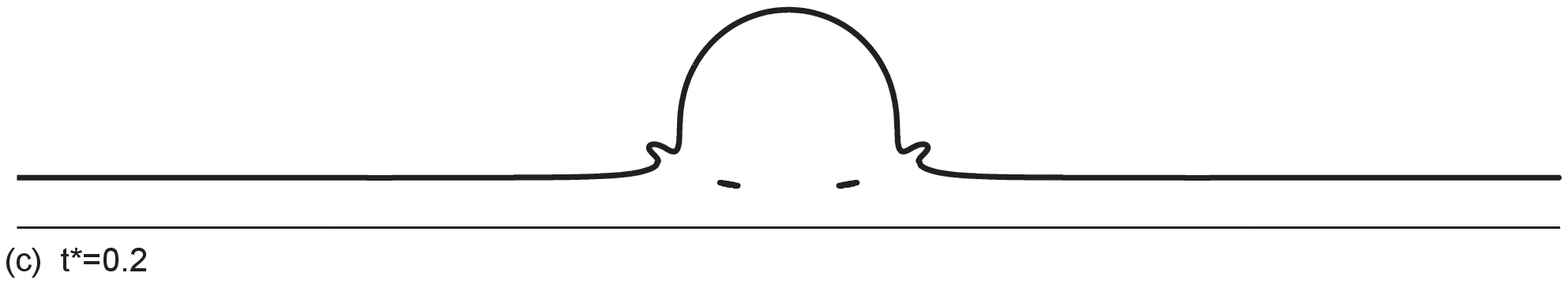}
\includegraphics[width=5.5in,height=1.4in]{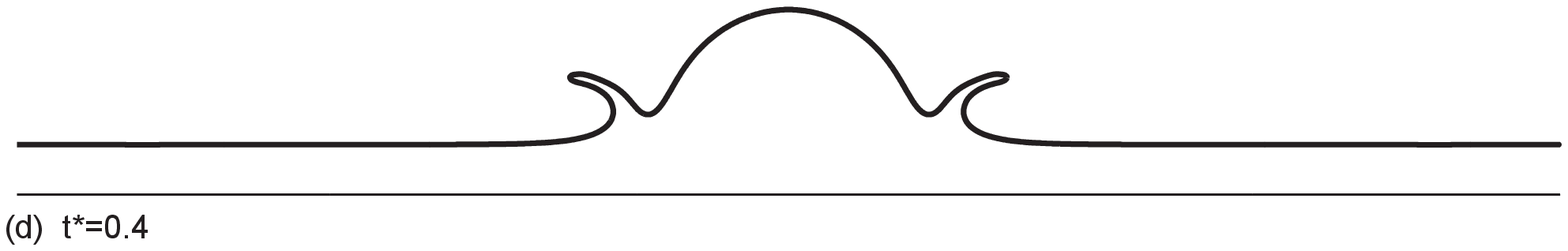}
\includegraphics[width=5.5in,height=1.4in]{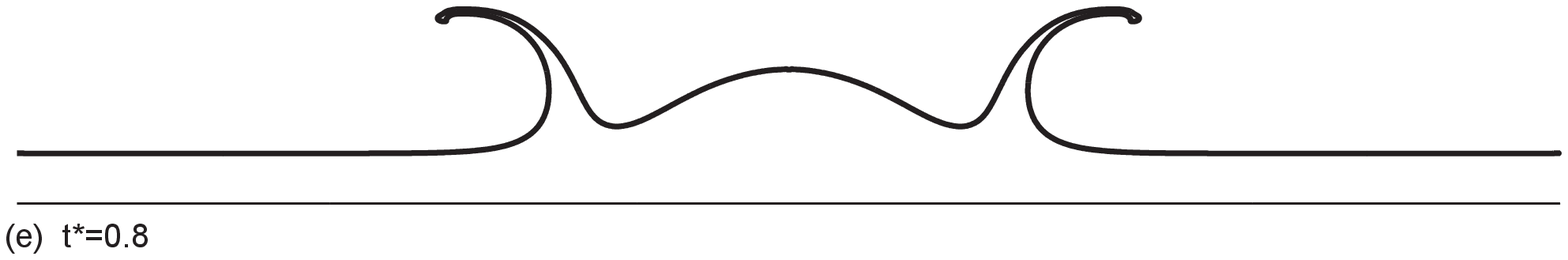}
\includegraphics[width=5.5in,height=1.4in]{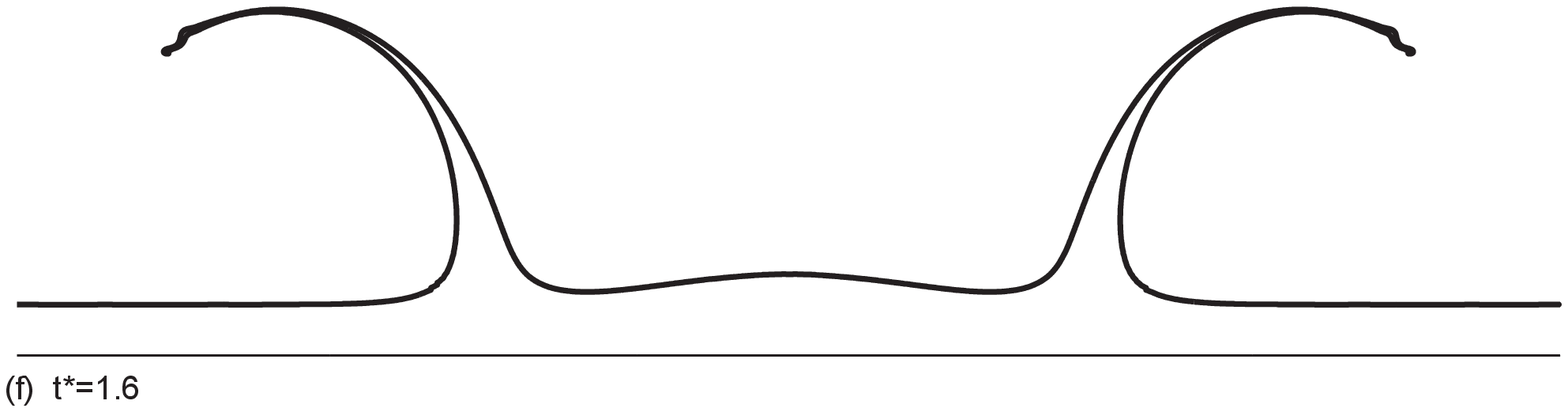}
 \tiny\caption{Snapshots of droplet impact on a thin liquid film with $Re=500$, $We=8000$ and $\rho_l/\rho_g=1000$. The
time instants $t^*$ have been normalized by the characteristic time
$D/U$.}
\end{figure}

\begin{figure}
\centering
\includegraphics[width=5.5in,height=1.4in]{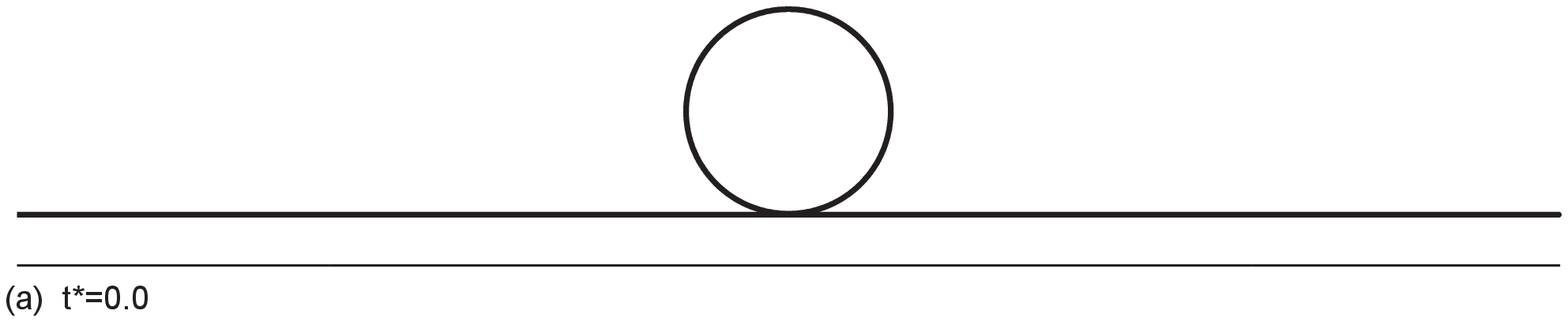}
\includegraphics[width=5.5in,height=1.4in]{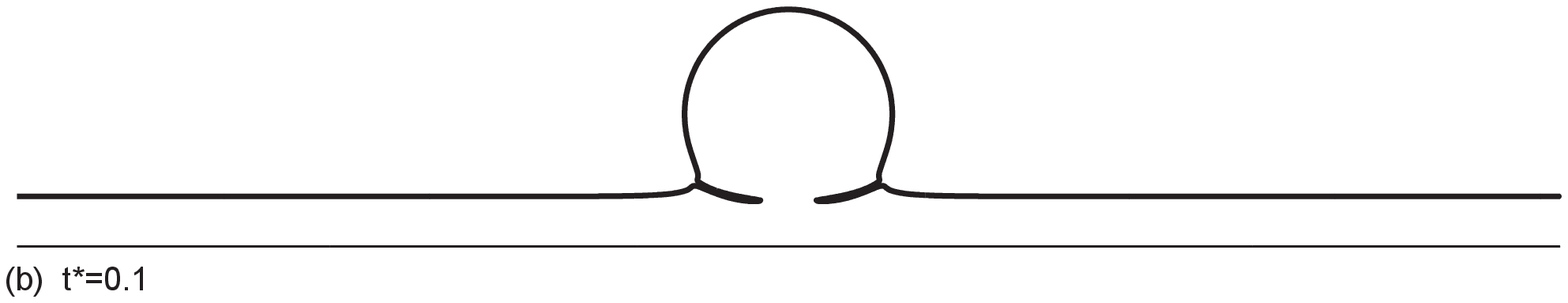}
\includegraphics[width=5.5in,height=1.4in]{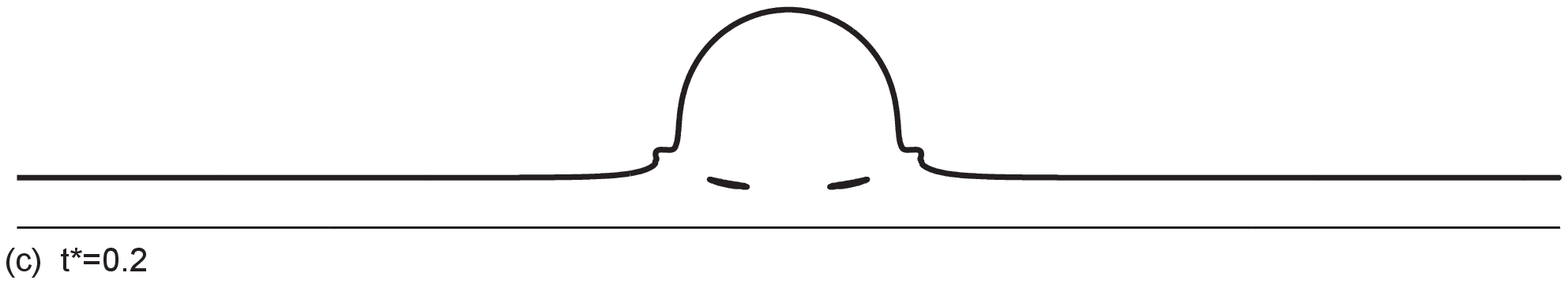}
\includegraphics[width=5.5in,height=1.4in]{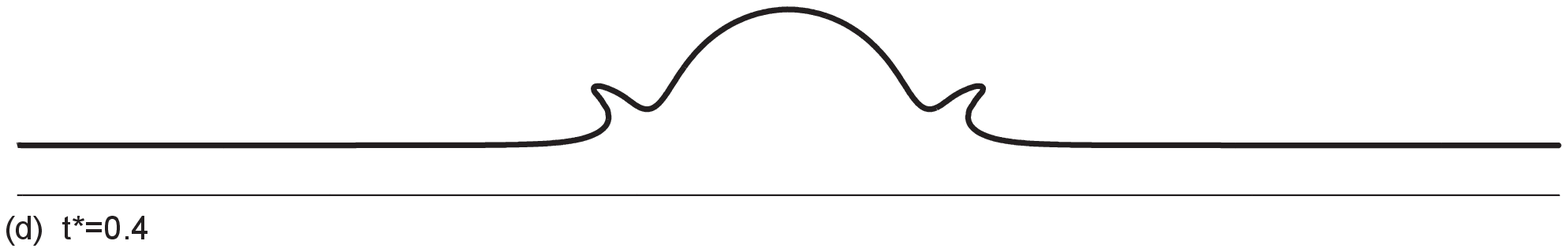}
\includegraphics[width=5.5in,height=1.4in]{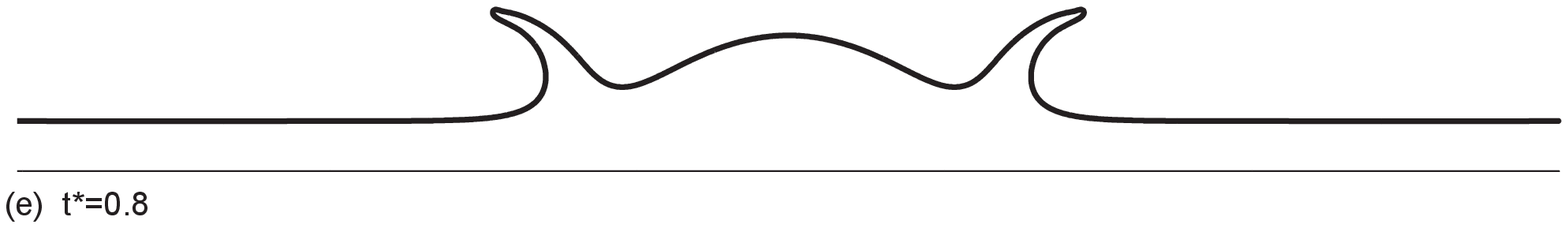}
\includegraphics[width=5.5in,height=1.4in]{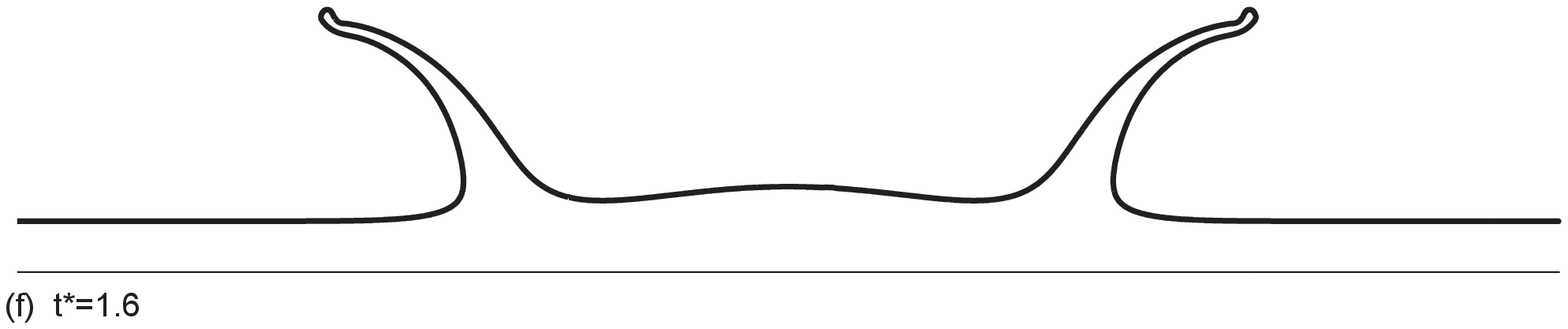}
 \tiny\caption{Snapshots of droplet impact on a thin liquid film with $Re=100$, $We=8000$ and $\rho_l/\rho_g=1000$. The
time instants $t^*$ have been normalized by the characteristic time
$D/U$.}
\end{figure}

\begin{figure}
\centering
\includegraphics[width=5.5in,height=1.4in]{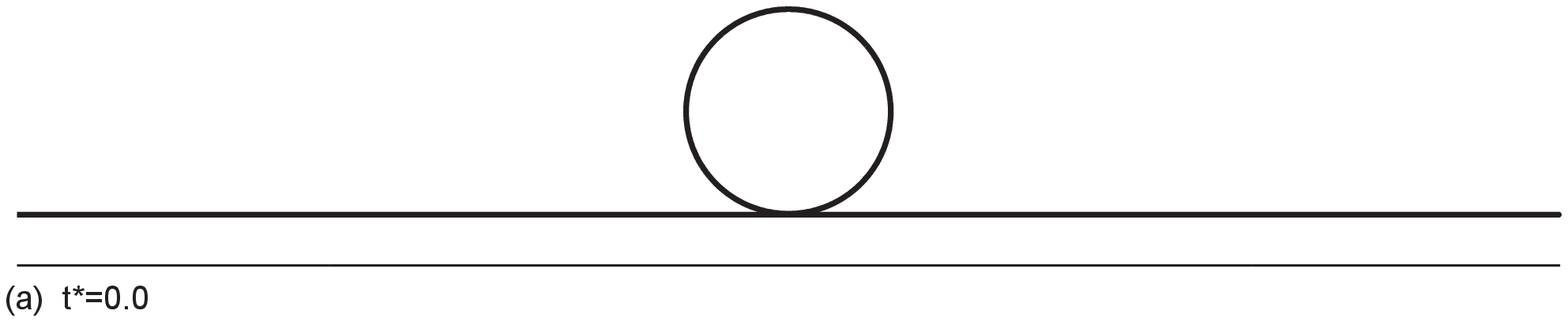}
\includegraphics[width=5.5in,height=1.4in]{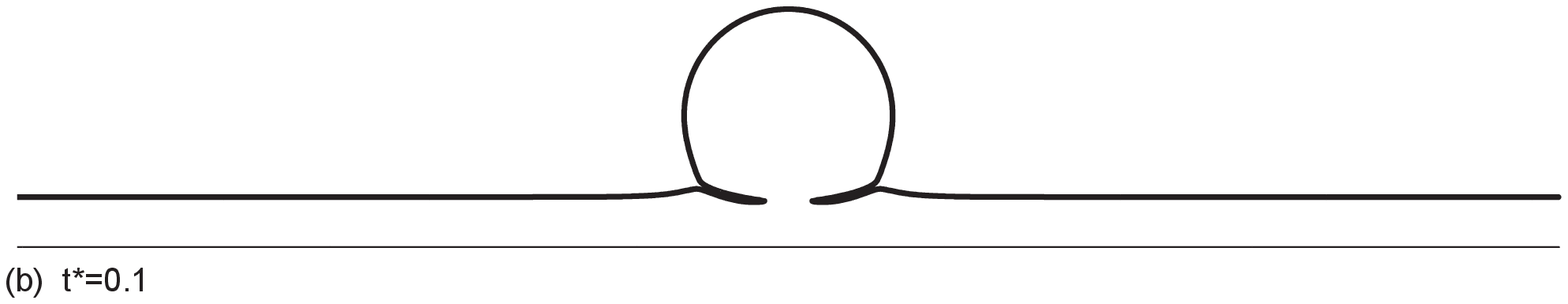}
\includegraphics[width=5.5in,height=1.4in]{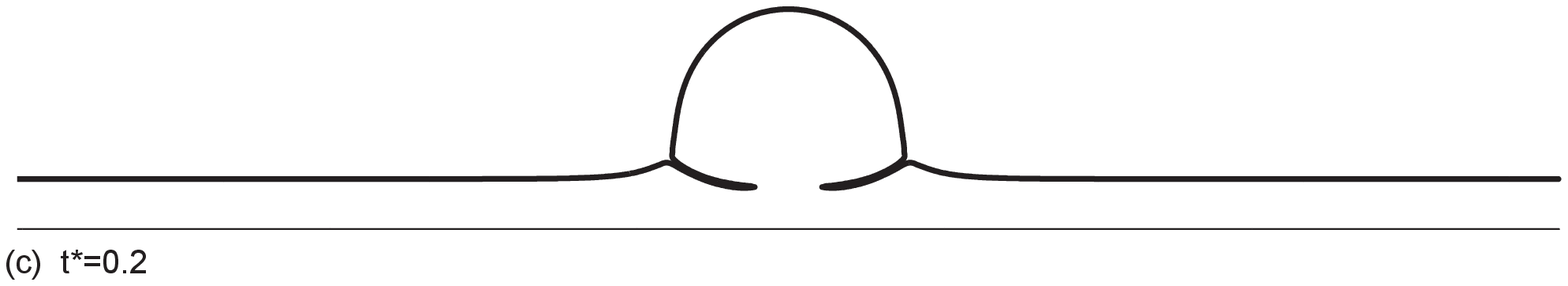}
\includegraphics[width=5.5in,height=1.4in]{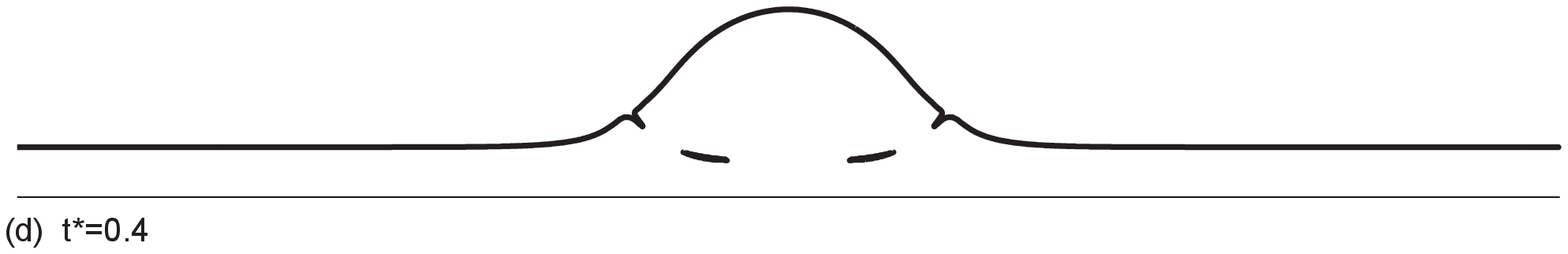}
\includegraphics[width=5.5in,height=1.4in]{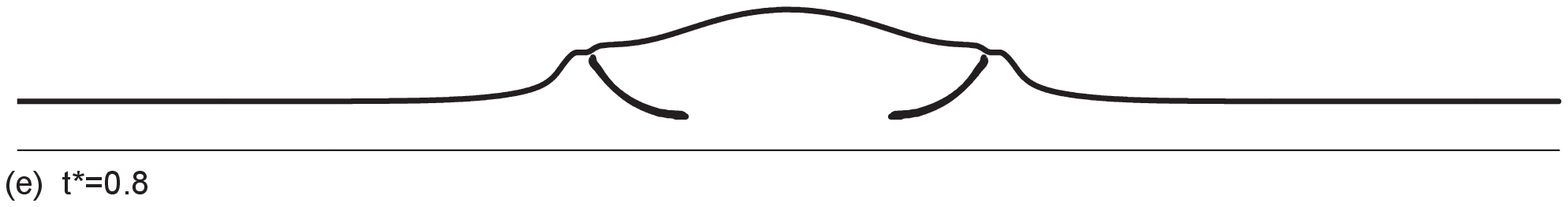}
\includegraphics[width=5.5in,height=1.4in]{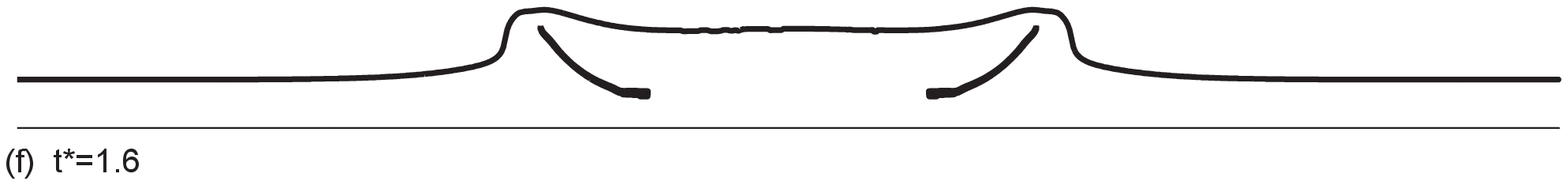}
 \tiny\caption{Snapshots of droplet impact on a thin liquid film with $Re=20$, $We=8000$ and $\rho_l/\rho_g=1000$. The
time instants $t^*$ have been normalized by the characteristic time
$D/U$.}
\end{figure}

The simulations are performed on a uniform computational mesh with
the size of $L \times H=1500\times 500$, as illustrated in Fig. 4. A
wetting liquid film with the height of $H_w=0.1H$ is initially
located at the bottom wall, and a circular droplet with the radius
($R$) of $100$ lattice units is just placed on the upper of the
liquid film. In the simulation, the distribution of the order
parameter can be initialized by
\begin{equation}
 \phi(x,y)= 0.5+0.5\tanh \frac{2\left(H_w-y\right)}{W},
\end{equation}
and also
\begin{equation}
\phi(x,y)=0.5+0.5\tanh
\frac{2\left[R-(x-0.5L)^2-(y-R-H_w)^2\right]}{W},~y>H_w,
\end{equation}
where $W$ is the interface thickness and is set to be 5. The
velocity field at initial time can be assigned by
\begin{equation}
(u,v)=\left\{
\begin{array}{ll}
(0,-\phi U),          ~y>H_w,    \\
(0,0),                ~~~~~ y\leq H_w, \\
\end{array}
\right.
\end{equation}
where $U$ is the impact velocity with a fixed value of $0.05$. The
periodic boundary conditions are applied at the left and right
boundaries, while the no-slip bounce back boundary condition is
imposed at the bottom wall and the open boundary condition is
utilized at the top boundary. Two major dimensionless parameters
governing droplet impact are the Reynolds number and the Weber
number, which are respectively defined by~\cite{Yarin}
\begin{equation}
Re=\frac{\rho_l D U}{\mu_l},
\end{equation}
and
\begin{equation}
We=\frac{\rho_l D U^2}{\sigma},
\end{equation}
where $D$ is the droplet diameter. The Weber number has been taken
fixed and equals to be $We=8000$, as commonly used in other studies
~\cite{Josseranda, Lee1, Coppola}. The density ratio of the liquid and gas
phases is set to $\rho_l/\rho_g=1000:1$. Three typical Reynolds numbers $Re=500$,
100, and 20 are considered in this work, which are derived by tuning
the kinematic viscosity of the liquid phase while keeping the gas
kinematic viscosity as a constant. With this strategy, it is found
that the lowest achievable liquid kinematic viscosity is 0.02 at the
largest $Re$ of 500, and the viscosity ratio $(\nu_g/\nu_l)$ at this
situation is 10, which is very close to that of a realistic
water-air two-phase system. With the driving of the impact velocity,
the system is released and the droplet will instantly impact onto
the underneath film. Here we mainly concentrate on the interfacial
dynamics and the variation law of the spreading radius versus time.
Figs. 5-7 depict typical scenic representations of droplet impact
process at three different Reynolds numbers of 20, 100 and 500,
where the time instants $t^*$ is the normalized time defined by
$t^*=tU/D$, $t$ is the iteration step. For high $Re$ of 500, the
droplet moves downward instantly with slight deformation at initial
stage, and some tiny bubble-ring entrapments are visible in the neck
connecting the droplet and film. The bubble entrapment phenomenon
has also been reported in the recent studies on the droplet impact
~\cite{JLee, Thoraval}. Then the droplet continues to spread, followed
by the formation of the ejecta sheet at the intersection region between
the droplet and liquid layer. The ejecta sheet grows into a splashing lamella
(known as crown in axisymmetric or 3D geometry) propagating radially with
increasing time and tends to bend at its end rim. The splashing
phenomenon is also observed in a moderate $Re$ of 100, as shown in
Fig. 6, while the extent is significantly reduced. This is ascribed
to the increasing frictional force between the liquid phase and its
ambient vapor phase at a larger viscosity, and then the interface
layer can be more stable. As the Reynolds number is lowered to a
small value of 20, we do not observe the splashing behavior of the
droplet in Fig. 7, as expected. The droplet only merge with the thin
liquid film, which evolves in a manner of the outward moving surface
wave. This process of droplet impact is oftentimes named as
deposition, which is in line with the results of the previous
studies~\cite{Lee1, Li2}.

We also conducted a quantitative study on the spreading radius,
which is a concerning physical quantity in droplet impact dynamics
~\cite{Yarin}. The previous researches~\cite{Josseranda, Coppola, Ba, Li2, Lee1}
have indicated that the growth of the spreading radius generally can be described
by the power law $r/D=C\sqrt{Ut/D}$, where $C$ is a coefficient that depends on the
flow geometry. For the axisymmetric or 3D modeling of the droplet
impact, the coefficient $C$ is found to be 1.1 by Josseranda and
Zaleskib~\cite{Josseranda}. While for the plane two-dimensional situation,
the scaled prefactor $C$ is found to be larger than 1.1, as reported in several
literatures~\cite{Lee1, Ba, Li1, Coppola}. Figure 8 shows the time variation of the numerically
predicted spreading radius by the present model. For a comparison,
the theoretical results of the fitting power formula is also
presented. The comparison between them shows a good agreement in
general, except for slight deviation at initial instants. In
addition, the spreading radius in the present simulation exhibits to
obey the power law $r/D=1.35\sqrt{Ut/D}$.

\begin{figure}
\centering
\includegraphics[width=4.0in,height=3.4in]{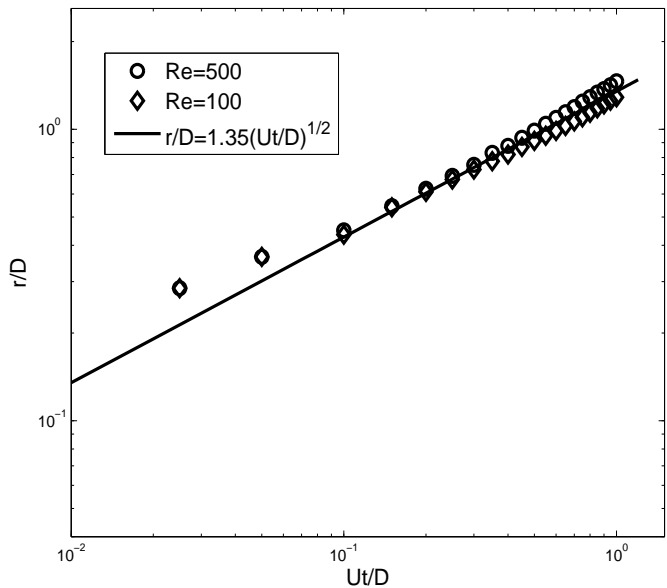}
 \tiny\caption{The numerically predicted spreading radius versus the dimensionless time. The solid line represents the theoretical power law.}
\end{figure}

\section{Summary}\label{sec: sum}
Numerical modeling of two-phase flows with large density ratios
is still a challenging task in the framework of LB approach.
In this paper, we propose a simple and accurate LB model for two-phase systems,
which is capable of simulating large-density-ratio flows. The proposed
LB model is based on the conservative phase-field equation, which involves a
lower-order diffusion term compared with the commonly used Cahn-Hilliard equation in
interface capturing. Therefore, the present model is expected to achieve a better numerical accuracy
and stability. In addition, a novel force distribution function is also
is elaborately designed in this model such that it contains only one nonlocal macroscopic quantity,
which is much simpler than the previous phase-field-based LB models~\cite{Lee1, Lee2, Zu, Liang1, Ren, Fakhari2}. The multi-scale
analysis also demonstrates that both the conservative Allen-Cahn equation and the incompressible
Navier-Stokes equations can be derived correctly from the present model.
To validate the present model, we firstly simulated two basic steady problems
of static droplet and layered Poiseuille flows, which have their own
analytical solutions. In the former test, it is found that
the present can accurately capture the density field distributions in the bulk regions
and also across the interface at the density ratio of 1000. In addition, it is also shown that
the present model can obtain relatively small spurious velocities in the LB community, with the
maximum magnitude of the order of $10^{-9}$. In the latter test, we simulated the
channel flow with a density ratio ranging from 10 to 1000, and also conducted
detailed comparisons with the previous Allen-Cahn based LB models~\cite{Ren, Fakhari2}. It is found that
the present model can obtain satisfactory results in the velocity predictions, and is also
more accurate than the previous LB models~\cite{Ren, Fakhari2}. Next, we consider two dynamic problems of
Spinodal decomposition and droplet impact on a thin liquid film with a large density ratio of 1000.
The phase separation process can be clearly observed in the system, which is in line with the expectation. The present model also
successfully reproduces the classical splashing phenomenon, and the predicted
spreading radius is found to exhibit the power law reported in the literature,
which provides a good validation of the present LB model in dealing with complex high-density-ratio two-phase flows.
Finally, we anticipate that our numerical method will be useful to scientific applications, such as
micro-fluidics, material science, and oil recovery industry.


\section*{Acknowledgments}
One of the authors (Hong Liang) gratefully thanks insightful discussions with Prof. Qing Li
in the study of droplet impact problem. This work is financially supported by the National Natural Science Foundation
of China(Grant Nos. 11602075 and 51576079).

\appendix

\section{Chapman-Enskog analysis of the present model}
The Chapman-Enskog analysis is now performed to demonstrate the consistency of the LB
evolution equation (15) with the incompressible Navier-Stokes equations. The moment conditions is first given
based on the expressions of the equilibrium and force distribution functions:
\begin{equation}
\begin{split}
\sum\limits_i {g_i^{eq}}=0, ~\sum\limits_i&
 c_{i\alpha}{g_i^{eq}}=\rho{u_\alpha},\\
\sum\limits_i c_{i\alpha}c_{i\beta}{g_i^{eq}}=\rho{u_\alpha}{u_\beta}+p\delta_{\alpha\beta},&~
\sum\limits_i{c}_{i\alpha}{c}_{i\beta}{c}_{i\gamma}{g_i^{eq}}=\rho{c_s^2}\Delta_{\alpha\beta\gamma\theta}{u_\theta},
\end{split}
\end{equation}
\begin{eqnarray}
 \sum\limits_i {G_i}=(1-\frac{1}{2\tau_g})u_\alpha\partial_\alpha \rho,~
 \sum\limits_i c_{i\alpha}{G_i}=(1-\frac{1}{2\tau_g}){F_\alpha},\\ \nonumber
\Lambda=:\sum\limits_i{{c}_{i\alpha}}{{c}_{i\beta}}{G_i}=(1-\frac{1}{2\tau_g})[{u_\alpha}\partial_\beta(\rho{c_s^2})
 +{u_\beta}\partial_\alpha(\rho{c_s^2})
 +({u_\gamma}\partial_\gamma \rho {c_s^2}){\delta_{\alpha\beta}}],
\end{eqnarray}
where $\delta_{\alpha\beta}$ is the Kronecker delta function, $\Delta_{\alpha \beta \gamma \theta}={\delta}_{\alpha\beta}
{\delta}_{\gamma\theta}+{\delta}_{\alpha\gamma}\delta_{\beta\theta}+
{\delta}_{\alpha\theta}{\delta}_{\beta\gamma}$. To derive the macroscopic equations, we expand the particle distribution function,
the time and space derivatives, and the force in consecutive scales of $\epsilon$,
\begin{subequations}
\begin{equation}
{{g}_i} = {g}_i^{(0)} + \epsilon {g}_i^{(1)} + {\epsilon^2}{g}_i^{(2)} + \cdot  \cdot  \cdot ,
\end{equation}
\begin{equation}
{\partial _t} = \epsilon {\partial _{t_1}} + {\epsilon ^2}{\partial
_{t_2}},~\partial_\alpha  = \epsilon {\partial _{1\alpha}},
\end{equation}
\begin{equation}
{F_\alpha}=\epsilon{F_\alpha^{(1)}},
\end{equation}
\end{subequations}
where $\epsilon$ is a small expansion parameter.
Applying the Taylor expansion to Eq. (15), and substituting Eq. (A3) into the expanded result,
we can obtain the following multi-scale equations,
\begin{subequations}
\begin{equation}
{\epsilon^0}:~{g}_i^{(0)} = {g}_i^{(eq)},
\end{equation}
\begin{equation}
{\epsilon ^1}:~{D_{1i}} {g}_i^{(0)} =  - {1 \over {{\tau_g\delta
_t}}}{g}_i^{(1)} + {G_i^{(1)}},
\end{equation}
\begin{equation}
{\epsilon ^2}:~{\partial _{{t_2}}}{g}_i^{(0)} + {D_{1i}}{g}_i^{(1)}
+ {{{\delta _t}} \over 2}D_{1i}^2{g}_i^{(0)} = - {1 \over
{{\tau_g\delta _t}}}{g}_i^{(2)},
\end{equation}
\end{subequations}
where ${D_{1i}}={\partial_{t_1}} + {c_{i\alpha}}\partial_{1\alpha}$.
The substitution of Eq. (A4b) into Eq. (A4c) yields
\begin{equation}
{\partial _{{t_2}}}{g}_i^{(0)} +
{D_{1i}}(1-\frac{1}{2\tau_g}){g}_i^{(1)} =-{1\over {{\tau_g\delta
_t}}}{g}_i^{(2)}-\frac{\delta_t}{2}{D_{1i}}{G_i^{(1)}}.
\end{equation}
Following Refs.~\cite{Liang1,Liang3}, the zero-order moment of ${g_i}$ can be defined as,
\begin{equation}
\sum\limits_k{g_k}=-\frac{\delta_t}{2}{u_\alpha}{\partial_\alpha}\rho.
\end{equation}
Applying the expansion formula (A3) to Eqs. (21a) and (A6), one can easily derive
\begin{equation}
\sum\limits_i{g_i^{(1)}}=-\frac{\delta_t}{2}u_\alpha{\partial_{1\alpha}}\rho,~
\sum\limits_i{g_i^{(n)}}=0,~(n\geq2),
\end{equation}
\begin{equation}
\sum\limits_i{c_{i\alpha}}{g_i^{(1)}}=-\frac{\delta_t}{2}{F_\alpha^{(1)}},
~\sum\limits_i{{c}_{i\alpha}}{g_i^{(n)}}=0,~(n\geq2).
\end{equation}
The recovered equations at ${\epsilon}$ scale can be obtained by summing Eq. (A4b) and Eq. (A4b)$\times$ ${c_{i\beta}}$ over $i$, respectively,
\begin{equation}
\partial_{1\alpha}{u_\alpha}=0,
\end{equation}
\begin{equation}
{\partial _{t_1}{(\rho
u_\beta)}}+\partial_{1\alpha}(\rho{u_\alpha}u_\beta+p\delta_{\alpha\beta})={F_\beta^{(1)}}.
\end{equation}
Similarly, the recovered equations at ${\epsilon ^2}$ scale can be derived from Eq. (A5),
\begin{equation}
\partial _{t_1}({-\frac{\delta_t}{2}
u_\alpha{\partial_{1\alpha}}\rho})+\partial_{1\alpha}(-\frac{\delta_t}{2}{F_\alpha^{(1)}})=-\frac{\delta_t}{2}[\partial
_{t_1}(u_\alpha{\partial_{1\alpha}}\rho)+
\partial_{1\alpha}{F_\alpha^{(1)}}]
\end{equation}
\begin{equation}
 {\partial _{t_2}{(\rho
{u_\beta})}}+(1-\frac{1}{2\tau_g})\partial_{1\alpha}
\Pi^{(1)}=-\frac{\delta_t}{2}\partial_{1\alpha}\Lambda^{(1)},
\end{equation}
where $\Pi^{(1)}=\sum\limits_i{c_{i\alpha}}{c_{i\beta}}g_i^{(1)}$ is
the first-order momentum flux tensor determined below, and $\Lambda =\epsilon\Lambda^{(1)}$. From Eq. (A4b), one can get
\begin{equation}
\begin{split}
\Pi^{(1)}&=\sum\limits_i{c_{i\alpha}}{c_{i\beta}}g_i^{(1)}=
-{\tau_g}{\delta_t}\sum\limits_i{c_{i\alpha}}{c_{i\beta}}\left[D_{1i}g_i^{(0)}-{G_i^{(1)}}\right]\\
&=-{\tau_g}{\delta_t}{c_s^2}\left[{\partial_{1\alpha}}({\rho}{u_\beta})
+{{\partial_{1\beta}}({\rho}{u_\alpha}})+(\partial_{1\gamma}
\rho{u_\gamma}){\delta_{\alpha\beta}}\right]+{\tau_g}{\delta_t}\Lambda^{(1)},
\end{split}
\end{equation}
where the terms of $O({\delta_t}{Ma}^2)$ have been neglected under the incompressible limit.
Substituting Eq. (A13) into Eq. (A12), one can simplify Eq. (A12) as
\begin{equation}
{\partial
_{t_2}{(\rho{u_\beta})}}-\partial_{1\alpha}\left[\nu{\rho}\left({\partial_{1\alpha}}{u_\beta}
+{{\partial_{1\beta}}{u_\alpha}}\right)\right]=0,
\end{equation}
where $\nu=c_s^2\delta_t(\tau_g-\frac{1}{2})$ is the kinematic viscosity.
Combining Eqs. (A9) and (A11) at $\epsilon$ and $\epsilon^2$ scales, together with Eqs. (A10) and (A14), we have
\begin{equation}
 \partial_\alpha{u_\alpha}=0,
\end{equation}
\begin{equation}
{\partial
_{t}{(\rho{u_\beta})}}+\partial_\alpha(\rho{u_\alpha}{u_\beta}+p{\delta_{\alpha\beta}})
=\partial_{1\alpha}\left[\nu{\rho}\left({\partial_{1\alpha}}{u_\beta}
+{{\partial_{1\beta}}{u_\alpha}}\right)\right]+{F_\beta},
\end{equation}
which clearly shows that the incompressible Navier-Stokes equations
can be exactly recovered from the present LB model.



\begin{thebibliography}{99}
\bibitem{Succi}
S. Succi, The Lattice Boltzmann Equation for Fluid Dynamics
and Beyond, Oxford University Press, Oxford, 2001.

\bibitem{Guo}
Z. L. Guo, C. Shu, Lattice Boltzmann method and its applications in
engineering, World Scientific Singapore, 2013.

\bibitem{Shi}
B. C. Shi and Z. L. Guo, Lattice Boltzmann model for nonlinear
convection-diffusion equations, Rev. E 79 (2009) 016701.

\bibitem{Chai}
 Z. H. Chai, B. C. Shi, Z. L. Guo, A multiple-relaxation-time lattice
Boltzmann model for general nonlinear anisotropic convection-diffusion
equations, J. Sci. Comput. 69 (2016) 355-390.

\bibitem{Gunstensen}
A. K. Gunstensen, D. H. Rothman, S. Zaleski, and G. Zanetti, Lattice
Boltzmann model of immiscible fluids, Phys. Rev. A 43 (1991) 4320.

\bibitem{Shan}
X. Shan, H. Chen, Lattice Boltzmann model for simulating flows with multiple phases and components, Phys. Rev. E 47 (1993) 1815.

\bibitem{Swift}
M. Swift, W. Osborn, and J. Yeomans, Lattice Boltzmann simulation of
nonideal fluids, Phys. Rev. Lett. 75 (1995) 830.

\bibitem{He}
X. He, S. Chen, and R. Zhang, A lattice Boltzmann scheme for incom-
pressible multiphase flow and its application in simulation of Rayleigh-
Taylor instability, J. Comput. Phys. 152 (1999) 642-663.

\bibitem{Lee}
T. Lee and L. Liu, Lattice Boltzmann simulations of micron-scale drop
impact on dry surfaces, J. Comput. Phys. 229 (2010) 8045-8063.

\bibitem{Liang1}
H. Liang, B. C. Shi, Z. L. Guo, Z. H. Chai, Phase-field-based multiple-
relaxation-time lattice Boltzmann model for incompressible multiphase
flows, Phys. Rev. E 89 (2014) 053320.

\bibitem{Liang2}
H. Liang, Z. H. Chai, B. C. Shi, Z. L. Guo, and T. Zhang, Phase-field-
based lattice Boltzmann model for axisymmetric multiphase flows, Phys.
Rev. E 90 (2014) 063311.

\bibitem{Liu}
H. Liu, Q. J. Kang, C. R. Leonardi et. al., Multiphase lattice Boltzmann
simulations for porous media applications, Comput. Geosci. 20 (2016) 777-
805

\bibitem{Li1}
Q. Li, H. K. Luo, Q. J. Kang, Y. L. He, Q. Chen, Q. Liu, Lattice
Boltzmann methods for multiphase flow and phase-change heat transfer,
Prog. Energy Combust. Sci. 52 (2016) 62-105.

\bibitem{Inamuro}
T. Inamuro, T. Ogata, S. Tajima, N. Konishi, A lattice Boltzmann method for incompressible
two-phase flows with large density differences, J. Comput. Phys. 198 (2004) 628.

\bibitem{Sethian}
J. A. Sethian, Level set methods and fast marching methods: evolving interfaces in computational geometry,
fluid mechanics, computer vision, and materials science, Cambridge university press, 1999.

\bibitem{Yuan}
P. Yuan, L. Schaefer, Equations of state in a lattice Boltzmann model, Phys. Fluids, 18 (2006) 042101.

\bibitem{Li2}
Q. Li, K. H. Luo, X. J. Li, Lattice Boltzmannmodeling of multiphase flows at large density ratio
with an improved pseudopotential model, Phys. Rev. E 87 (2013) 053301.

\bibitem{Ba}
Y. Ba, H. Liu, Q. Li, Q. Kang, J. Sun, Multiple-relaxation-time color-gradient lattice Boltzmann model
for simulating two-phase flows with high density ratio, Phys. Rev. E 94 (2016) 023310.

\bibitem{Jacqmin}
D. Jacqmin, Calculation of two-phase Navier-Stokes flows using phase-field modeling, J. Comput. Phys. 155 (1999) 96-127.

\bibitem{Zheng}
H. W. Zheng, C. Shu, Y. T. Chew, A lattice Boltzmann model for multiphase flows with large density ratio,
J. Comput. Phys. 218 (2006) 353.

\bibitem{Fakhari1}
A. Fakhari, M.H. Rahimian, Phase-field modeling by the method of lattice Boltzmann equations, Phys. Rev. E 81 (2010) 036707.

\bibitem{Lee1}
T. Lee, C L. Lin, A stable discretization of the lattice Boltzmann equation for simulation of incompressible
two-phase flows at high density ratio, J. Comput. Phys. 206 (2005) 16-47.

\bibitem{Lee2}
T. Lee, L. Liu, Lattice Boltzmann simulations of micron-scale drop impact on dry surfaces, J. Comput. Phys. 229 (2010) 8045.

\bibitem{Lou}
Q. Lou, Z. L. Guo, B. C. Shi, Effects of force discretization on mass conservation in lattice Boltzmann
equation for two-phase flows, Europhys. Lett. 99 (2012) 64005.

\bibitem{Zu}
Y. Q. Zu, S. He, Phase-field-based lattice Boltzmann model for incompressible binary fluid systems with
density and viscosity contrasts, Phys. Rev. E 87 (2013) 043301.

\bibitem{Wang}
Y. Wang, C. Shu, H. B. Huang, C. T. Teo, Multiphase lattice Boltzmann flux solver for incompressible
multiphase flows with large density ratio, J. Comput. Phys. 280 (2015) 404-423.

\bibitem{Ren}
F. Ren, B. W. Song, M. C. Sukop, and H. B. Hu, Improved lattice Boltzmann modeling of binary flow
based on the conservative Allen-Cahn equation, Phys. Rev. E 94 (2016) 023311.

\bibitem{Fakhari2}
A. Fakhari, D. Bolster, Diffuse interface modeling of three-phase contact line dynamics on
curved boundaries: A lattice Boltzmann model for large density and viscosity ratios, J. Comput. Phys. 334 (2017) 620-638.

\bibitem{HWang}
H. L. Wang, Z. H. Chai, B. C. Shi, and H. Liang, Comparative study of the lattice Boltzmann
models for Allen-Cahn and Cahn-Hilliard equations, Phys. Rev. E 94 (2016) 033304.

\bibitem{Sun}
Y. Sun, C. Beckermann, Sharp interface tracking using the phase-field equation, J. Comput. Phys. 220 (2007) 626-653.

\bibitem{Chiu}
P. H. Chiu, Y. T. Lin, A conservative phase field method for solving incompressible two-phase flows,
J. Comput. Phys. 230 (2011) 185-204.

\bibitem{Qian}
Y. H. Qian, D. d¡¯Humires, and P. Lallemand, Lattice BGK models for Navier-Stokes equation, Europhys. Lett. 17 (1992) 479-484.

\bibitem{Ginzburg}
I. Ginzburg, F. Verhaeghe, D. d¡¯Humieres, Two-relaxation-time lattice Boltzmann scheme: About parametrization,
velocity, pressure and mixed boundary conditions, Commun. Comput. Phys. 3 (2008) 427-478.

\bibitem{Lallemand}
P. Lallemand, L. S. Luo, Theory of the lattice Boltzmann method: dispersion, dissipation,
isotropy, Galilean invariance, and stability, Phys. Rev. E 61 (2000) 6546.

\bibitem{Unverdi}
S. O. Unverdi, G. Tryggvason, A front-tracking method for viscous,
incompressible, multi-fluid flows, J. Comput. Phys. 100 (1992) 25-37.

\bibitem{Kim}
J. Kim, A continuous surface tension force formulation for diffuse-interface models, J. Comput. Phys. 204 (2005) 784-804.

\bibitem{Liang3}
H. Liang, B. C. Shi, and Z. H. Chai, Lattice Boltzmann modeling of three-phase incompressible flows, Phys. Rev. E 93 (2016) 013308.

\bibitem{Wei1}
Y. K. Wei, Z. D. Wang, H. S. Dou, Y. H. Qian, A novel two-dimensional coupled lattice Boltzmann model for incompressible
flow in application of turbulence Rayleigh-Taylor instability, Comput. Fluids 156 (2017) 97-102.

\bibitem{Guo1}
Z. L. Guo, C. G. Zheng, and B. C. Shi, Discrete lattice effects on the
forcing term in the lattice Boltzmann method, Phys. Rev. E 65 (2002) 046308.

\bibitem{Wei2}
Y. K. Wei, Z. D. Wang, J. F. Yang, H. S. Dou, Y. H Qian, A simple
lattice Boltzmann model for turbulence Rayleigh-B¨¦nard thermal
convection, Comput. Fluids 118 (2015) 167-171.

\bibitem{Anderson}
D. M. Anderson, G. B. McFadden, and A. A. Wheeler, Diffuse-interface
methods in fluid mechanics, Annu. Rev. Fluid Mech. 30 (1998) 139-165.

\bibitem{Guo2}
Z. L. Guo, C. G. Zheng, and B. C. Shi, Force imbalance in lattice Boltzmann equation for
two-phase flows, Phys. Rev. E 83 (2011) 036707.

\bibitem{Yu}
Z. Yu, L. S. Fan, Multirelaxation-time interaction-potential-based lattice Boltzmann model for two-phase flow,
Phys. Rev. E 82 (2010) 046708.

\bibitem{Huang}
H. B. Huang, X. Y. Lu, Relative permeabilities and coupling effects in steady-state gas-liquid flow in porous media:
A lattice Boltzmann study, Phys. Fluids 21 (2009) 092104.

\bibitem{Liang4}
H. Liang, B. C. Shi, Z. H. Chai, An efficient phase-field-based multiple-relaxation-time lattice Boltzmann model
for three-dimensional multiphase flows, Comput. Math. Appl. 73 (2017) 1524-1538.

\bibitem{Cahn}
J. W. Cahn, Phase separation by spinodal decomposition in isotropic systems, J. Chem. Phys. 42 (1965) 93-99.

\bibitem{Gan}
Y. B. Gan, A. G. Xu, G. C. Zhang, and S. Succi, Discrete Boltzmann modeling of multiphase flows:
hydrodynamic and thermodynamic non-equilibrium effects, Soft Matter 11 (2015) 5336-5345.

\bibitem{Yarin}
A. L. Yarin, Drop impact dynamics: splashing, spreading, receding,
bouncing..., Annu. Rev. Fluid Mech. 38 (2006) 159-192.

\bibitem{Coppola}
G. Coppola, G. Rocco, and L. Luca, Insights on the impact of a plane drop on a thin liquid film, Phys. Fluids 23 (2011) 022105.

\bibitem{JLee}
J. S. Lee, B. M. Weon, J. H. Je, and K. Fezzaa, How does an air film
evolve into a bubble during drop impact?, Phys. Rev. Lett. 109
(2012) 204501.

\bibitem{Thoraval}
M. J. Thoraval, K. Takehara, T. G. Etoh and S. T. Thoroddsen, Drop impact entrapment of bubble rings,
J. Fluid Mech. 724 (2013) 234-258.

\bibitem{Josseranda}
C. Josseranda and S. Zaleskib, Droplet splashing on a thin liquid film, Phys. Fluids 15 (2003) 1650.


\end{thebibliography}
\end{document}